\documentclass[9pt]{extarticle}
\pdfoutput=1
\ifdefined\pdfminorversion\pdfminorversion=7\fi
\usepackage{microtype}
\usepackage[utf8]{inputenc}
\usepackage[T1]{fontenc}
\usepackage{url}
\usepackage{amssymb}
\usepackage{amsmath,amsfonts}
\usepackage{pifont}
\usepackage[square,numbers]{natbib}
\usepackage{colortbl}
\usepackage{multirow}
\usepackage{amsthm}
\usepackage{mathtools}
\usepackage{booktabs}
\usepackage{tabularx,longtable,array}
\newcolumntype{Y}{>{\raggedright\arraybackslash}X}
\usepackage{textcomp}
\usepackage[table,xcdraw]{xcolor}
\usepackage{fancyhdr}

\definecolor{cai_primary}{HTML}{4C9A99}
\definecolor{cai_secondary}{HTML}{307FE2}
\definecolor{cai_accent}{HTML}{1D8348}
\definecolor{cai_dark}{HTML}{3F4444}
\definecolor{human_color}{HTML}{173C47}

\usepackage{graphicx}
\usepackage{subcaption}
\usepackage{float}
\usepackage{placeins}
\usepackage{stfloats}
\usepackage{hyperref}
\hypersetup{
    pdftitle={Browser Immune System (BIS): Innate and Adaptive Threat Detection in the Browser with a Cybersecurity Transformer},
    pdfauthor={Víctor Mayoral-Vilches, Eduardo Worrel},
    pdfsubject={In-browser threat detection with a cybersecurity transformer on WebGPU},
    colorlinks=true, urlcolor=cai_secondary, linkcolor=cai_secondary, filecolor=cai_accent, citecolor=cai_secondary,
}
\makeatletter
\g@addto@macro{\UrlBreaks}{\do\/\do\-\do\_\do\.\do\=\do\?\do\&}
\makeatother

\usepackage{tikz}
\usetikzlibrary{arrows.meta,positioning,calc,shadows,shapes.geometric,fit,backgrounds}
\usepackage{pgfplots}
\pgfplotsset{compat=1.16}
\definecolor{fig_teal}{HTML}{2E6260}
\definecolor{fig_mini}{HTML}{4C9A99}
\definecolor{fig_base}{HTML}{B8D8D8}
\definecolor{fig_s1}{HTML}{0CA30C}
\definecolor{fig_s2}{HTML}{1F4E8C}
\definecolor{fig_compose}{HTML}{2E6260}
\definecolor{fig_thinking}{HTML}{B8641B}
\definecolor{fig_control}{HTML}{5B6875}
\definecolor{fig_ink}{HTML}{16202A}
\definecolor{fig_rule}{HTML}{D6DCE2}
\definecolor{fig_warning}{HTML}{8A1C1C}
\definecolor{fig_frontier}{HTML}{C0392B}
\definecolor{fig_surface}{HTML}{F7F8FA}
\definecolor{fig_process}{HTML}{7A4FB5}
\definecolor{fig_time}{HTML}{6B7884}
\definecolor{fig_link}{HTML}{307FE2}

\makeatletter
\def\DejaVuSansMono@scale{0.9}
\makeatother
\IfFileExists{T1DejaVuSansCondensed-TLF.fd}
  {\newcommand{\figfonts}{\renewcommand{\sfdefault}{DejaVuSansCondensed-TLF}\renewcommand{\ttdefault}{DejaVuSansMono-TLF}}}
  {\newcommand{\figfonts}{}}%
\tikzset{
  lane/.style={rounded corners=6pt, fill=fig_surface, draw=fig_rule, line width=.5pt},
  lanet/.style={font=\sffamily\bfseries\fontsize{6.2}{7}\selectfont, text=fig_control, anchor=west, inner sep=0pt},
  card/.style={rounded corners=3pt, draw=#1, fill=#1!10, line width=.7pt, inner sep=0pt,
               drop shadow={shadow xshift=.6pt, shadow yshift=-.6pt, opacity=.10, fill=fig_ink}},
  plain/.style={rounded corners=3pt, draw=fig_rule, fill=white, line width=.5pt, inner sep=0pt},
  chip/.style={rounded corners=4pt, draw=fig_teal, fill=fig_teal!9, line width=.9pt, inner sep=0pt,
               drop shadow={shadow xshift=.8pt, shadow yshift=-.8pt, opacity=.12, fill=fig_ink}},
  pin/.style={fill=fig_teal, inner sep=0pt, minimum width=2.6pt, minimum height=1.4pt},
  t/.style={font=\sffamily\fontsize{5.6}{6.6}\selectfont, text=fig_ink, inner sep=0pt},
  tt/.style={font=\sffamily\bfseries\fontsize{6.3}{7.3}\selectfont, text=fig_ink, inner sep=0pt},
  ts/.style={font=\sffamily\fontsize{5.3}{6.3}\selectfont, text=fig_control, inner sep=0pt},
  tx/.style={font=\sffamily\fontsize{4.9}{5.6}\selectfont, text=fig_control, inner sep=0pt},
  tm/.style={font=\ttfamily\fontsize{5}{6}\selectfont, text=fig_control, inner sep=0pt},
  tmi/.style={font=\ttfamily\fontsize{5}{6}\selectfont, text=fig_ink, inner sep=0pt},
  flow1/.style={-{Latex[length=2.1mm,width=1.5mm]}, draw=fig_s1, line width=.9pt, rounded corners=3pt},
  flow2/.style={-{Latex[length=2.1mm,width=1.5mm]}, draw=fig_s2, line width=.9pt, dashed, rounded corners=3pt},
  bus/.style={-{Latex[length=1.9mm,width=1.4mm]}, draw=fig_control, line width=.75pt, rounded corners=3pt},
  thin/.style={draw=fig_control!70, line width=.45pt},
  lab/.style={font=\sffamily\fontsize{5}{6}\selectfont, text=fig_control, inner sep=1.2pt, fill=white, rounded corners=1pt},
  pill/.style={rounded corners=5pt, inner sep=0pt, minimum height=11pt, font=\ttfamily\fontsize{5.2}{6}\selectfont},
  tbadge/.style={circle, fill=#1, text=white, inner sep=0pt, minimum size=8.2pt, font=\sffamily\bfseries\fontsize{4.8}{5}\selectfont},
}
\newcommand{\shield}[5]{%
  \begin{scope}[shift={(#1,#2)}, scale=#3]
    \fill[#4] (-.5,.42) -- (0,.55) -- (.5,.42) -- (.5,-.05) .. controls (.5,-.45) and (0,-.62) .. (0,-.62)
      .. controls (0,-.62) and (-.5,-.45) .. (-.5,-.05) -- cycle;
    \node[font=\sffamily\bfseries\fontsize{4.5}{5}\selectfont, text=white] at (0,-.02) {#5};
  \end{scope}}

\newcommand{\bistable}{\small\setlength{\tabcolsep}{6pt}\renewcommand{\arraystretch}{1.2}\arrayrulecolor{fig_teal!60}}
\newcommand{\tabhead}{\rowcolor{fig_teal!12}}
\newcommand{\tabalt}{\rowcolor{fig_teal!4}}
\newcommand{\lat}[1]{\textcolor{fig_time}{#1}}
\DeclareRobustCommand{\badge}[2][fig_s1]{\tikz[baseline=(b.base)]\node[circle,fill=#1,text=white,inner sep=0pt,minimum size=1.05em,font=\sffamily\bfseries\tiny](b){#2};}

\definecolor{hf_face}{HTML}{FFD21E}
\definecolor{hf_hand}{HTML}{FF9D00}
\definecolor{hf_finger}{HTML}{E8890C}
\definecolor{hf_ink}{HTML}{3A3B45}
\DeclareRobustCommand{\hfemoji}{\texorpdfstring{%
  \begin{tikzpicture}[baseline=-0.30ex,x=1.18ex,y=1.18ex,line join=round,line cap=round]
    \foreach \s in {-1,1}{%
      \begin{scope}[shift={({\s*1.00},-0.18)},rotate=\s*-18]
        \fill[hf_hand] (0,0) ellipse [x radius=0.40, y radius=0.50];
        \foreach \d in {-0.17,0,0.17}{\draw[hf_finger,line width=0.055ex] (\d,0.40) -- (\d,-0.26);}
      \end{scope}}
    \fill[hf_face] (0,0) circle [radius=0.98];
    \fill[hf_ink] (-0.38,0.22) circle [radius=0.155];
    \fill[hf_ink] ( 0.38,0.22) circle [radius=0.155];
    \draw[hf_ink,line width=0.15ex] (-0.40,-0.26) .. controls (-0.20,-0.60) and (0.20,-0.60) .. (0.40,-0.26);
  \end{tikzpicture}}{}}

\newcommand{\aliasmicrourl}{https://huggingface.co/aliasrobotics/alias3-micro-1026}
\DeclareRobustCommand{\aliasmicro}{\texorpdfstring{\href{\aliasmicrourl}{\textcolor{fig_teal}{\texttt{alias3-micro}}}}{alias3-micro}}
\newcommand{\aliasweburl}{https://huggingface.co/aliasrobotics/alias3-web}
\DeclareRobustCommand{\aliasweb}{\texorpdfstring{\href{\aliasweburl}{\textcolor{fig_teal}{\texttt{alias-web}}}}{alias-web}}
\DeclareRobustCommand{\bisbench}{\texorpdfstring{\texttt{BISBENCH-8}}{BISBENCH-8}}
\DeclareRobustCommand{\basemodel}{\texttt{Qwen3.5-4B}}
\newcommand{\sone}{System~1}
\newcommand{\stwo}{System~2}

\usepackage{enumitem}
\usepackage{listings}
\usepackage{geometry}
\renewcommand{\headrulewidth}{0.4pt}
\renewcommand{\footrulewidth}{0.4pt}
\renewcommand{\headrule}{\hbox to\headwidth{\color{cai_primary}\leaders\hrule height \headrulewidth\hfill}}
\renewcommand{\footrule}{\hbox to\headwidth{\color{human_color}\leaders\hrule height \footrulewidth\hfill}}
\usepackage{caption,setspace}
\usepackage{titlesec}
\titleformat{\section}{\normalfont\Large\bfseries\color{cai_primary}}{\thesection}{1em}{}[\titlerule]
\titleformat{\subsection}{\normalfont\large\bfseries\color{human_color}}{\thesubsection}{1em}{}
\titleformat{\subsubsection}{\normalfont\normalsize\bfseries\color{cai_dark}}{\thesubsubsection}{1em}{}

\usepackage{authblk}

\renewcommand\Affilfont{\small\normalfont}
\makeatletter
\renewcommand\AB@affilsepx{\\\protect\Affilfont}
\let\orig@maketitle\maketitle
\renewcommand{\maketitle}{\orig@maketitle\vspace{-1.5em}{\color{cai_primary!30}\hrule height 0.5pt}\vspace{1em}}
\makeatother

\makeatletter
\renewenvironment{abstract}{\small\noindent\ignorespaces}{\par}
\makeatother

\begin{document}

\title{\LARGE\textcolor{cai_primary}{\textbf{Browser Immune System (BIS)}}\\[0.25em]
\large\textcolor{cai_dark}{Innate and Adaptive Threat Detection in the Browser with a Cybersecurity Transformer}}

\author[1,$\dagger$]{V\'ictor Mayoral-Vilches}
\author[2]{Eduardo Worrel}
\affil[1]{\small Alias Robotics, Vitoria-Gasteiz, \'Alava, Spain}
\affil[2]{\small Independent researcher}

\date{}
\twocolumn[
\vspace*{-2.4em}
\maketitle
\vspace{0.3em}
\begin{abstract}
Security decisions about a web page are best made where the page is: in the browser, before anything leaves the machine. We present the \textbf{Browser Immune System (BIS)}, a browser extension that screens navigation with a cybersecurity transformer running entirely on the user's GPU, and \aliasweb{}, the model that powers it: a $4$\,B-parameter cybersecurity transformer that starts from the dual-regime model of Mayoral-Vilches et al.~\cite{mayoralvilches2026transformer}, extends its public release \aliasmicro{} with a context-category sensor, and is quantised to 4-bit weights and compiled to a $2.44$\,GB graph for WebGPU. \aliasweb{} is the first in-browser realisation of that model's typed decision mode, \sone{}, with its official pieces, the marker-token mapping, the complete pointer head and the final normalisation that the exported graph omits; the preceding edge study~\cite{worrel2026system1edge} had to recover the head from a damaged release file and write the markers as plain text. It answers a typed security question in about $150$\,ms on a laptop GPU, in the decision direction of the bf16 reference where we compared them, and the same weights generate explanations at $10.5$ tokens/s on demand, $6$ to $39\times$ the cost of a decision. BIS treats the model as an immune system: an \emph{innate} response, a battery of \sone{} questions with one question per threat class selected by the kind of observation, runs on every page; an \emph{adaptive} response, \stwo{} text, explains a verdict only when the user asks. To measure it we introduce \bisbench{}, eight synthetic sites: six execute real malicious behaviour (credential exfiltration, keylogging, mining, an obfuscated dropper, scareware, fingerprinting) against a local collector that proves each signal was present, and two are adversarial controls. BIS returns the expected page verdict on $7/8$ sites, with both controls green and a class-level micro-F1 of $0.71$, scans a page in $0.9$ to $6.4$\,s, and its verdicts are identical across independent samples and across Chrome and Firefox, the latter $5.9\times$ slower. Single-variable diagnostics on the exact observations trace both benchmark errors to what the model was shown (a self-describing comment in a fixture, a comparison of host names) and, with a first pass over eleven real public pages, show that two of the seven questions answer yes on ordinary scripts and login forms: specificity on the real web is the open problem. We report every error, the twenty-five engineering obstacles between a research checkpoint and a working extension, and what the evidence does not yet support.

\end{abstract}
\vspace{0.5em}
{\centering\small\hfemoji~\textbf{Open weights:}~\url{\aliasweburl}\par}
\vspace{0.45em}
\noindent\resizebox{\textwidth}{!}{%
\figfonts
\begin{tikzpicture}[x=1cm,y=1cm]
\draw[lane] (0,1.95) rectangle (5.2,9.35);
\draw[lane] (5.35,1.95) rectangle (6.45,9.35);
\draw[lane] (6.6,1.95) rectangle (12.5,9.35);
\draw[lane] (12.65,1.95) rectangle (17.2,9.35);
\node[lanet] at (0.18,9.12) {BROWSER $\cdot$ the page};
\node[lanet] at (5.5,9.12) {BUS};
\node[lanet] at (6.78,9.12) {ENGINE TAB $\cdot$ \texttt{alias-web} on the user's GPU};
\node[lanet] at (12.83,9.12) {USER INTERFACE};

\draw[plain, drop shadow={shadow xshift=.6pt, shadow yshift=-.6pt, opacity=.1, fill=fig_ink}] (0.2,2.15) rectangle (3.55,8.8);
\fill[fig_rule!55, rounded corners=3pt] (0.2,8.35) rectangle (3.55,8.8);
\fill[fig_rule!55] (0.2,8.35) rectangle (3.55,8.55);
\foreach \x in {0.42,0.6,0.78} \fill[fig_control!55] (\x,8.575) circle (.055);
\draw[plain] (0.95,8.43) rectangle (3.0,8.71);
\node[tm] at (1.975,8.57) {127.0.0.1:8930};
\shield{3.22}{8.6}{.28}{fig_warning}{2}
\node[font=\sffamily\bfseries\fontsize{6.8}{7}\selectfont, text=fig_ink, anchor=west] at (0.42,8.03) {Banco Seguro};
\node[font=\sffamily\fontsize{5.1}{5.6}\selectfont, text=fig_ink, anchor=west] at (0.42,7.76) {Verificaci\'on requerida de su cuenta};
\fill[fig_thinking!9, rounded corners=1.5pt] (0.42,6.62) rectangle (3.33,7.62);
\fill[fig_thinking!70] (0.42,6.62) rectangle (0.47,7.62);
\node[font=\sffamily\fontsize{5}{6.1}\selectfont, text=fig_ink, text width=2.66cm, anchor=north west, align=left] at (0.55,7.57) {Estimado cliente: hemos detectado acceso inusual. Verifique su cuenta en 24 horas o ser\'a suspendida.};
\draw[rounded corners=2pt, fill=fig_surface, draw=fig_control!60, line width=.5pt] (0.42,4.42) rectangle (3.33,6.5);
\draw[plain] (0.57,6.08) rectangle (3.18,6.34); \node[tm, anchor=west] at (0.65,6.21) {usuario};
\draw[plain] (0.57,5.74) rectangle (3.18,6.0); \node[tm, anchor=west] at (0.65,5.87) {\textbullet\textbullet\textbullet\textbullet\textbullet\textbullet\textbullet};
\draw[plain] (0.57,5.4) rectangle (3.18,5.66); \node[tm, anchor=west] at (0.65,5.53) {tel\'efono};
\fill[fig_s2!75, rounded corners=1.5pt] (0.57,5.02) rectangle (2.2,5.28); \node[font=\sffamily\bfseries\fontsize{5}{5}\selectfont, text=white] at (1.385,5.15) {Verificar ahora};
\node[font=\ttfamily\fontsize{4.9}{5}\selectfont, text=fig_warning, anchor=west] at (0.57,4.82) {action=:8931/collect/login};
\node[font=\sffamily\fontsize{5.1}{5}\selectfont, text=fig_warning, anchor=west] at (0.57,4.6) {same host, other port};
\node[tx, text=fig_control!80, anchor=west] at (0.42,4.2) {\textcopyright\ Banco Seguro 2026};
\draw[rounded corners=2pt, draw=fig_rule, dashed, line width=.5pt] (0.42,3.25) rectangle (3.33,3.95);
\node[font=\sffamily\itshape\fontsize{5.1}{5.6}\selectfont, text=fig_control!75] at (1.875,3.6) {no inline script on this page};
\draw[rounded corners=2pt, draw=fig_rule, dashed, line width=.5pt] (0.42,2.4) rectangle (3.33,3.1);
\node[font=\sffamily\itshape\fontsize{5.1}{5.6}\selectfont, text=fig_control!75] at (1.875,2.75) {no frame, no external script};
\node[font=\sffamily\bfseries\fontsize{5.8}{6}\selectfont, text=fig_ink, anchor=east] (obsl) at (5.06,9.12) {observations}; \node[tbadge=fig_control, anchor=east] at ([xshift=-2.5pt]obsl.west) {5};
\foreach \y/\lbl in {7.14/page, 5.46/form}{
  \draw[thin] (3.33,\y) -- (3.69,\y);
  \draw[rounded corners=2pt, fill=white, draw=fig_control, line width=.55pt] (3.69,\y-.17) rectangle (5.08,\y+.17);
  \node[tmi] at (4.385,\y) {\lbl};
}
\foreach \y/\lbl/\src in {3.6/script-inline/3.6, 2.98/script-src/2.75, 2.56/iframe/2.75}{
  \draw[fig_rule, line width=.5pt, dashed] (3.33,\src) -- (3.69,\y);
  \draw[rounded corners=2pt, fill=fig_surface, draw=fig_rule, dashed, line width=.5pt] (3.69,\y-.17) rectangle (5.08,\y+.17);
  \node[tm, text=fig_control!70] at (4.385,\y) {\lbl};
}
\node[ts, text width=1.5cm, align=center] at (4.4,4.5) {content.js packages $\leq$14 observations of $\leq$1\,400 chars; it never decides};

\draw[rounded corners=4pt, fill=fig_control!14, draw=fig_control, line width=.7pt] (5.5,2.3) rectangle (6.3,8.55);
\node[rotate=90, font=\sffamily\bfseries\fontsize{5.6}{6.6}\selectfont, text=fig_control] at (5.9,5.45) {service worker $\cdot$ chrome.storage bus};
\foreach \y in {7.14,5.46} \draw[thin] (5.08,\y) -- (5.5,5.45);
\node[lab] at (5.9,8.78) {bis-scan};
\draw[bus] (6.3,5.45) -- (6.75,5.45);
\node[lab, anchor=south] at (6.52,5.52) {req};

\node[chip, minimum width=5.5cm, minimum height=1.5cm, anchor=south west] (chipn) at (6.8,7.25) {};
\foreach \y in {7.5,7.8,8.1,8.4} {\node[pin] at (6.78,\y) {}; \node[pin] at (12.32,\y) {};}
\node[tbadge=fig_teal] at (7.12,8.52) {1}; \node[tt, anchor=west] at (7.3,8.52) {\texttt{alias-web}\ $\cdot$ 4\,B transformer $\cdot$ int4 WebGPU};
\node[ts, anchor=west] at (7.3,8.22) {2.44 GB $\cdot$ 32 layers $\cdot$ loaded once in 2.9--7 s};
\node[tbadge=fig_teal] at (7.12,7.92) {2}; \node[ts, anchor=west] at (7.3,7.92) {official readout: marker ids $\cdot$ \texttt{head.bin} $\cdot$ \texttt{norm.bin}};
\node[ts, anchor=west] at (7.3,7.6) {on-device inference $\cdot$ nothing leaves the machine};
\draw[bus] (9.55,7.25) -- (9.55,7.15);
\draw[rounded corners=4pt, fill=fig_s1!8, draw=fig_s1, line width=.8pt] (6.8,4.05) rectangle (12.3,7.12);
\node[tbadge=fig_s1] at (7.09,6.93) {3}; \node[tt, anchor=west, text=fig_s1!70!black] at (7.27,6.93) {innate response $\cdot$ System 1 battery};
\node[ts, anchor=west] at (6.95,6.66) {\texttt{POLICY(kind)} $\to$ one directed yes/no question per class};
\foreach \i/\name/\val/\w/\asked in {0/phish-\\ing/0.00/0/1, 1/scare-\\ware/0.999/.54/1, 2/key-\\logger/1.00/.54/1, 3/obfus-\\cation/{}/0/0, 4/drop-\\per/{}/0/0, 5/miner/{}/0/0, 6/finger-\\print/{}/0/0}{
  \pgfmathsetmacro{\xa}{6.95+\i*0.77}
  \ifnum\asked=1
    \draw[rounded corners=2pt, fill=white, draw=fig_s1!70, line width=.6pt] (\xa,5.22) rectangle (\xa+.7,6.35);
    \node[font=\sffamily\fontsize{5}{5.4}\selectfont, text=fig_ink, align=center] at (\xa+.35,6.12) {\name};
    \fill[fig_rule, rounded corners=1pt] (\xa+.08,5.5) rectangle (\xa+.62,5.62);
    \ifdim\w pt>0pt \fill[fig_s1, rounded corners=1pt] (\xa+.08,5.5) rectangle (\xa+.08+\w,5.62);\fi
    \node[tm, text=fig_ink] at (\xa+.35,5.35) {\val};
  \else
    \draw[rounded corners=2pt, fill=fig_surface, draw=fig_rule, dashed, line width=.5pt] (\xa,5.22) rectangle (\xa+.7,6.35);
    \node[font=\sffamily\fontsize{5}{5.4}\selectfont, text=fig_control, align=center] at (\xa+.35,6.12) {\name};
    \node[font=\sffamily\fontsize{4.8}{5.3}\selectfont, text=fig_control, align=center] at (\xa+.35,5.5) {not\\asked};
  \fi
}
\node[ts, anchor=west] at (6.95,4.78) {verdict: max\,$P$ = 1.00 $\geq$ 0.8 \ $\to$};
\node[pill, fill=fig_warning, text=white, minimum width=24pt, font=\sffamily\bfseries\fontsize{5.4}{6}\selectfont] at (10.28,4.78) {RED};
\node[ts, anchor=west] at (10.81,4.78) {in 927 ms};
\node[ts, anchor=west] at (6.95,4.4) {4 questions, one prefill each $\cdot$ \texttt{FLAG} 0.5 $\cdot$ \texttt{STRONG} 0.8};
\draw[rounded corners=4pt, fill=fig_s2!8, draw=fig_s2, dashed, line width=.8pt] (6.8,2.2) rectangle (12.3,3.75);
\node[tbadge=fig_s2] at (7.09,3.55) {4}; \node[tt, anchor=west, text=fig_s2] at (7.27,3.55) {adaptive response $\cdot$ System 2, on request};
\node[ts, anchor=west] at (6.95,3.22) {prompt: the findings + ``explain this risk briefly''};
\node[ts, anchor=west] at (6.95,2.92) {forced \texttt{</think>} prefix $\cdot$ 6--9 tokens/s $\cdot$ $\leq$80 tokens $\cdot$ $\approx$4 s};
\node[font=\sffamily\itshape\fontsize{5.5}{6}\selectfont, text=fig_ink, anchor=west] at (6.95,2.5) {``El sitio presenta un riesgo cr\'itico al simular\ldots''\,\rule[-.1ex]{.5pt}{1.3ex}};
\draw[flow1] (12.3,5.65) -- (12.9,5.65);
\node[lab, anchor=south] at (12.6,5.72) {verdict};
\draw[flow1] (12.3,6.9) -- (12.575,6.9) -- (12.575,8.52) -- (12.85,8.52);
\node[lab, rotate=90] at (12.575,7.7) {level};
\draw[flow2, {Latex[length=2.1mm,width=1.5mm]}-{Latex[length=2.1mm,width=1.5mm]}] (12.3,3.0) to[out=0,in=180] (12.95,4.17);
\node[lab, text=fig_s2] at (12.62,3.45) {why?};

\fill[fig_rule!55, rounded corners=3pt] (12.85,8.3) rectangle (17.0,8.75);
\draw[plain] (12.97,8.38) rectangle (15.0,8.66); \node[tm] at (13.985,8.52) {127.0.0.1:8930};
\shield{16.6}{8.54}{.28}{fig_warning}{2}
\node[ts, anchor=east] at (16.95,8.1) {badge: \ding{55}\,2 findings $\cdot$ zero cost};
\node[plain, minimum width=4.1cm, minimum height=5.7cm, anchor=south west, drop shadow={shadow xshift=.8pt, shadow yshift=-.8pt, opacity=.12, fill=fig_ink}] at (12.9,2.2) {};
\draw[fig_rule, line width=.5pt] (12.9,7.45) -- (17.0,7.45);
\node[tbadge=fig_s2!40!fig_ink] at (13.19,7.68) {6}; \shield{13.55}{7.69}{.26}{fig_teal}{}
\node[font=\sffamily\bfseries\fontsize{5.9}{6}\selectfont, text=fig_ink, anchor=west] at (13.76,7.68) {Browser Immune System};
\draw[rounded corners=2.5pt, fill=white, draw=fig_rule, line width=.5pt] (13.05,6.84) rectangle (16.85,7.36);
\draw[fig_warning, line width=2pt, line cap=round] (13.1,6.93) -- (13.1,7.27);
\fill[fig_warning!14] (13.44,7.1) circle (.17); \node[font=\sffamily\bfseries\fontsize{5}{5}\selectfont, text=fig_warning] at (13.44,7.1) {\ding{55}};
\node[font=\sffamily\bfseries\fontsize{6.2}{6}\selectfont, text=fig_warning, anchor=west] at (13.72,7.215) {Amenaza detectada};
\node[tm, anchor=west] at (13.72,6.965) {4 preguntas S1 $\cdot$ 927 ms};
\foreach \x/\w/\lbl in {13.05/1.35/1$\times$ formulario, 14.5/1.0/1$\times$ p\'agina, 15.6/.6/2 obs}{
  \draw[rounded corners=4pt, fill=white, draw=fig_rule, line width=.5pt] (\x,6.47) rectangle (\x+\w,6.72);
  \node[tm, font=\ttfamily\fontsize{4.7}{5}\selectfont] at (\x+\w/2,6.595) {\lbl};
}
\foreach \y/\name/\pct/\w/\ev in {5.55/Scareware/99{,}9\%/3.4/p\'agina, 4.55/Keylogger/100{,}0\%/3.45/formulario}{
  \draw[rounded corners=2.5pt, fill=white, draw=fig_rule, line width=.5pt] (13.05,\y) rectangle (16.85,\y+.85);
  \draw[fig_warning, line width=2pt, line cap=round] (13.1,\y+.1) -- (13.1,\y+.75);
  \node[font=\sffamily\bfseries\fontsize{5.9}{6}\selectfont, text=fig_ink, anchor=west] at (13.28,\y+.66) {\name};
  \node[font=\ttfamily\bfseries\fontsize{5.2}{6}\selectfont, text=fig_warning, anchor=east] at (16.72,\y+.66) {\pct};
  \fill[fig_rule, rounded corners=1pt] (13.28,\y+.4) rectangle (16.72,\y+.5);
  \fill[fig_warning, rounded corners=1pt] (13.28,\y+.4) rectangle (13.28+\w,\y+.5);
  \node[font=\sffamily\fontsize{5.1}{5}\selectfont, text=fig_s2, anchor=west] at (13.28,\y+.19) {$\triangleright$ ver evidencia (\ev)};
}
\fill[fig_s2, rounded corners=3pt] (13.05,3.95) rectangle (16.85,4.38);
\node[font=\sffamily\bfseries\fontsize{5.7}{6}\selectfont, text=white, anchor=east] at (14.92,4.165) {?`Por qu\'e?};
\fill[white, opacity=.22, rounded corners=1.5pt] (15.04,4.065) rectangle (15.94,4.265);
\node[font=\ttfamily\bfseries\fontsize{4.5}{5}\selectfont, text=white] at (15.49,4.165) {SYSTEM 2};
\draw[rounded corners=2.5pt, fill=white, draw=fig_rule, line width=.5pt] (13.05,2.42) rectangle (16.85,3.8);
\node[font=\sffamily\fontsize{5.4}{6.3}\selectfont, text=fig_ink, text width=3.55cm, anchor=north west, align=left] at (13.17,3.73) {El sitio presenta un riesgo cr\'itico al simular una verificaci\'on bancaria para instalar software malicioso y registrar sus teclas.};
\node[tm, anchor=west] at (13.17,2.56) {System 2 $\cdot$ 9 tok/s $\cdot$ 25 tokens};
\node[ts] at (14.95,2.05) {100\,\% local $\cdot$ nothing leaves the machine};

\node[tbadge=fig_thinking] at (0.19,1.5) {7}; \node[tt, anchor=west] at (0.38,1.5) {\texttt{BISBENCH-8}};
\node[ts, anchor=west, text width=2.8cm, align=left] at (0.05,1.02) {6 real but harmless threats\\and 2 adversarial controls};
\foreach \i/\name/\verd/\col/\mark in {0/clean/green/fig_s1/51, 1/adheavy/green/fig_s1/51, 2/phishing/red/fig_warning/51, 3/obfuscated/red/fig_warning/51, 4/keylogger/red/fig_warning/51, 5/miner/green/fig_warning/55, 6/scareware/red/fig_warning/51, 7/fingerprint/red/fig_warning/51}{
  \pgfmathsetmacro{\xp}{2.95+\i*1.78}
  \ifnum\i=5
    \draw[rounded corners=5pt, fill=white, draw=fig_warning, dashed, line width=.6pt] (\xp,0.78) rectangle (\xp+1.66,1.25);
    \node[font=\ttfamily\fontsize{5.2}{6}\selectfont, text=fig_warning] at (\xp+.83,1.015) {miner \ding{55} missed};
  \else
    \draw[rounded corners=5pt, fill=\col!12, draw=\col, line width=.6pt] (\xp,0.78) rectangle (\xp+1.66,1.25);
    \node[font=\ttfamily\fontsize{5.2}{6}\selectfont, text=\col!80!black] at (\xp+.83,1.015) {\name\ \ding{\mark}};
  \fi
}
\node[ts] at (8.6,0.5) {\ding{51} = verdict as expected $\cdot$ 7/8 verdicts $\cdot$ 2/2 controls green $\cdot$ micro-F1 0.71 $\cdot$ 16/16 identical across samples $\cdot$ 8/8 across Chrome and Firefox};
\draw[fig_s1, line width=.9pt] (0.1,0.17) -- (0.6,0.17); \node[ts, anchor=west] at (0.68,0.17) {System 1 flow (innate)};
\draw[fig_s2, line width=.9pt, dashed] (3.0,0.17) -- (3.5,0.17); \node[ts, anchor=west] at (3.58,0.17) {System 2 flow (adaptive, on request)};
\draw[fig_control, line width=.75pt] (7.45,0.17) -- (7.95,0.17); \node[ts, anchor=west] at (8.03,0.17) {transport (storage bus)};
\node[tbadge=fig_control] at (10.75,0.17) {n}; \node[ts, anchor=west] at (10.93,0.17) {numbered as in Table~\ref{tab:elements}};
\end{tikzpicture}}
\par
\captionsetup{hypcap=false}

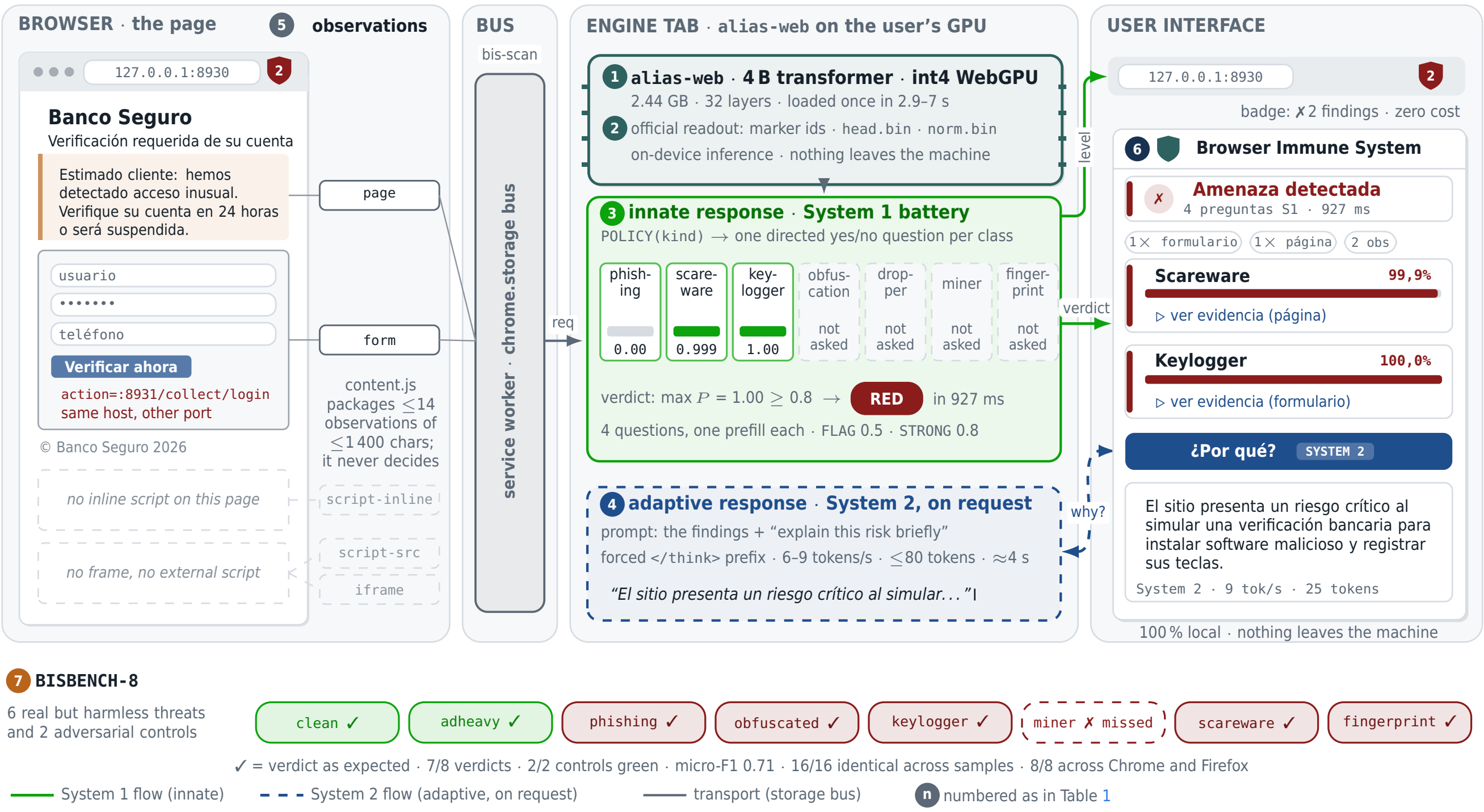
\captionof{figure}{\textbf{The Browser Immune System, as deployed, on the phishing site of \bisbench{}.} \emph{Left:} a bank notice with a deadline and a login form that posts to the collector is reduced by a content script to typed \emph{observations} that carry no detection logic \badge[fig_control]{5}, two on this page, and relayed over a storage bus. \emph{Centre:} \aliasweb{} \badge[fig_teal]{1}, a $4$\,B cybersecurity transformer compiled to a $2.44$\,GB 4-bit graph on the user's GPU, with its official readout pieces \badge[fig_teal]{2}, answers the \emph{innate} response \badge{3}: one directed yes/no question per threat class, selected by the kind of observation, each one forward pass with no token decoded; the receptors show the real probabilities of this page (scareware $0.999$ on its text, keylogger $1.00$ on its form, phishing $0.00$ because the observation reports the collector as the same domain; Section~\ref{sec:bench-errors} explains the last two), and the page verdict is the maximum. The \emph{adaptive} response \badge[fig_s2]{4}, \stwo{} text from the same weights, runs only when the user asks. \emph{Right:} the badge and the popup \badge[fig_s2!40!fig_ink]{6}, with the confidence and the evidence of each finding. \emph{Bottom:} the \bisbench{} scoreboard \badge[fig_thinking]{7}.}
\label{fig:overview}
]
\renewcommand{\thefootnote}{$\dagger$}
\setcounter{footnote}{0}
\footnotetext{Corresponding author: \texttt{victor@aliasrobotics.com}}
\renewcommand{\thefootnote}{\arabic{footnote}}
\setcounter{footnote}{0}

\section{Introduction}\label{sec:introduction}
The material that a security check must read is often the material that must not leave the user's machine: the source of the page being visited, the form about to receive a password, the script that just started running. Hosted detection services solve this by sending digests, URLs or features to a server, which is fast and scalable but moves the decision, and part of the evidence, off the device~\cite{google_safebrowsing,whittaker2010largescale}. The alternative, a detector that runs in the browser, has until now meant hand-built feature pipelines specialised to one threat at a time: static classifiers for malicious JavaScript~\cite{curtsinger2011zozzle,canali2011prophiler}, emulation for drive-by downloads~\cite{cova2010detection,rieck2010cujo}, content features for phishing~\cite{zhang2007cantina}, instrumentation for cryptojacking~\cite{konoth2018minesweeper,kharraz2019outguard} and fingerprinting~\cite{iqbal2021fpinspector}. Each is a separate engineering effort with its own feature set and its own model.

The cybersecurity transformer of Mayoral-Vilches et al.~\cite{mayoralvilches2026transformer} offers a different instrument. It is a compact model with two inference regimes: its \sone{} mode answers a \emph{typed} security question (is this a credential form posting to another domain, does this script contain a packed payload that is decoded at runtime) in a single forward pass through a small pointer head, decoding no tokens, while its \stwo{} mode is an ordinary generative agent sharing the same weights. One model, asked the right question, covers many threat classes; the feature engineering of the classic detectors becomes question engineering. The question we set out to answer is whether that instrument can run \emph{inside} the browser, on the user's GPU, fast enough to screen every page visited, and whether the result is a useful protection rather than a demonstration.

Two pieces of work made the question concrete. That paper measured \sone{} in bf16 on a server GPU and left quantisation and on-device execution to future work. Worrel~\cite{worrel2026system1edge} then published builds of its public release, \aliasmicro{}, for four browser runtimes, compared each with the original on $172$ decision questions, and found no detectable accuracy loss in the 4-bit builds at that sample size, while they change a measurable share of individual answers; but the marker-token mapping and the complete decision head were not public, so that study ran a recovered head on a plain-text approximation of the \sone{} interface (Section~\ref{sec:prior}). This paper starts from both: it closes that gap and builds on what the study established.

\paragraph{Contributions.}
\begin{enumerate}[leftmargin=*,nosep]
  \item \textbf{\aliasweb{}, a cybersecurity transformer for the browser} (Section~\ref{sec:model}): \aliasmicro{} with a fused decision adapter and a context-category sensor adapter, compiled to a $2.44$\,GB 4-bit ONNX graph for WebGPU and served with the model's official pieces: the marker-to-token mapping, the complete $1.3$\,M-parameter pointer head, and the final RMSNorm that the exported graph omits and whose absence saturates the head. \sone{} answers in $152$\,ms on a laptop GPU and agrees in decision direction with the PyTorch bf16 reference on the prompt compared; the deployment is the first to run the typed decision mode as designed outside a server.
  \item \textbf{Both inference regimes from one in-browser weight set} (Section~\ref{sec:model-dual}): \stwo{} generation at $10.5$ tokens/s from the same loaded graph, producing correct diagnoses and fixes, at $6$--$39\times$ the cost of a \sone{} decision on the same snippet; the measurement that justifies spending generation only where the fast path has already decided something is wrong.
  \item \textbf{The Browser Immune System} (Section~\ref{sec:bis}): an extension architecture in which the page is reduced to typed observations without detection heuristics, a battery of \sone{} questions selected by observation kind forms the innate response, and \stwo{} forms the adaptive one; a three-step interface (badge, popup with evidence, explanation on request); and a port to Firefox with a one-file difference, validated to return identical verdicts.
  \item \textbf{\bisbench{}} (Section~\ref{sec:bench}): a benchmark of eight synthetic sites, six whose malicious behaviour really executes (against a local collector that logs every callback) but harms nothing and two adversarial controls, with published labels, formal metrics and a reproducible harness. BIS scores $7/8$ page verdicts and micro-F1 $0.71$ at class level, with both controls green; its verdicts are identical across samples and across browsers. Single-variable diagnostics on the exact observations localise both errors in what the model was shown and, with a probe on eleven real public pages, mark the limit of the current battery: two of its seven questions are not specific enough for the real web (Sections~\ref{sec:bench-errors} and~\ref{sec:bench-live}).
  \item \textbf{A reproducible engineering record} (Section~\ref{sec:discussion}, Appendix~\ref{app:engineering}): ten modifications and twenty-five obstacles between the research checkpoint and the working extension, two of which generalise beyond this model (the trained form of a readout must be honoured exactly, and exported hidden states must be checked for the normalisation the head expects), together with the correction to the prior work's timing figure that its own research archive records.
\end{enumerate}

\begin{table*}[t]
\centering\bistable
\caption{\textbf{What BIS adds to a browser, and what each element buys.} Every payoff is a measurement of Sections~\ref{sec:model} to~\ref{sec:bench}, taken on the deployed build.}
\label{tab:elements}
\begin{tabularx}{\textwidth}{@{}c>{\raggedright\arraybackslash}p{.14\textwidth}>{\raggedright\arraybackslash}X>{\raggedright\arraybackslash}X>{\raggedright\arraybackslash}p{.07\textwidth}@{}}
\toprule
\tabhead & \textbf{Element} & \textbf{What it is} & \textbf{What it buys} & \textbf{Where} \\
\midrule
\badge[fig_teal]{1} & \aliasweb{} & \aliasmicro{} with fused decision and context-sensor adapters, compiled to a 4-bit ONNX graph ($2.44$\,GB) for WebGPU. & A cybersecurity transformer that loads in $2.3$\,s from a local server and runs entirely on the user's GPU; $152$\,ms per typed decision, $10.5$ tokens/s when it writes. & \S\ref{sec:model} \\
\tabalt \badge[fig_teal]{2} & Official \sone{} pieces & The marker-to-token mapping, the complete pointer head and the final RMSNorm applied in JavaScript before the head. & The head reads what it was trained on: without the norm it saturates ($\mathrm{e}^{-52}$); with it, the 4-bit graph answers the validation battery as expected and agrees in direction with the bf16 reference on the prompt compared. & \S\ref{sec:model-pieces} \\
\badge{3} & Innate response & A battery of yes/no \sone{} questions, one per threat class, chosen by the kind of observation (page, form, inline script, external script, iframe). & Seven threat classes from one model and one prefill per question; a page of the benchmark costs $4$--$9$ questions and $0.9$--$6.4$\,s; the directed question lifts the XSS probability from $0.040$ to $0.822$. & \S\ref{sec:bis-battery} \\
\tabalt \badge[fig_s2]{4} & Adaptive response & \stwo{} generation from the same weights, run only when the user asks why. & Correct plain-language diagnoses (DOMPurify for the XSS, parameterised queries for the SQL injection) at $6$--$11$ tokens/s, paid only on positives. & \S\ref{sec:model-dual}, \S\ref{sec:bis-ux} \\
\badge[fig_control]{5} & Observe, relay, decide & A content script that packages the DOM without heuristics, a service worker that is only a bus, an engine tab that owns the model. & Every error is attributable to the model or to the observation, never to a hidden rule; the same code runs in Chrome and Firefox with one manifest change and identical verdicts. & \S\ref{sec:bis-arch} \\
\tabalt \badge[fig_control]{6} & Three-step interface & Badge (no click), popup with per-class confidence and fold-out evidence (one click), explanation (one more). & The user sees a verdict at zero cost and can audit the exact observation that triggered it. & \S\ref{sec:bis-ux} \\
\badge[fig_thinking]{7} & \bisbench{} & Eight sites: six with behaviourally real, harmless malware verified by a local collector and two adversarial controls; labels and formal metrics. & $7/8$ page verdicts, both controls green, micro-F1 $0.71$; $16/16$ identical verdicts across samples, $8/8$ across browsers; a harness that reruns in minutes and localises an error by changing one variable. & \S\ref{sec:bench} \\
\bottomrule
\end{tabularx}
\end{table*}

Table~\ref{tab:elements} sets each element of BIS beside what it buys. Everything measured here runs on one laptop (Apple M3 Max, $128$\,GB) in Chrome and Firefox Nightly; the weights are fetched from a local server and no byte of page material leaves the machine (Section~\ref{sec:model-privacy}). The paper is organised as follows: Section~\ref{sec:related} places the work, Section~\ref{sec:prior} summarises the edge study it builds on, Sections~\ref{sec:model} to~\ref{sec:bench} present the model, the extension and the benchmark, and Section~\ref{sec:discussion} states what the evidence supports and what it does not.

\section{Related Work}\label{sec:related}
\paragraph{Detecting malicious web content.} Two decades of work detect malicious pages with specialised classifiers over hand-built features. ZOZZLE classifies JavaScript in the browser from abstract-syntax-tree features with a Bayesian model~\cite{curtsinger2011zozzle}; Prophiler filters pages at scale from HTML, JavaScript and URL features~\cite{canali2011prophiler}; JSAND~\cite{cova2010detection} and Cujo~\cite{rieck2010cujo} detect drive-by downloads by emulation and by static-dynamic analysis. Phishing detection moved from content heuristics~\cite{zhang2007cantina} to large-scale classifiers behind browser blocklists~\cite{whittaker2010largescale,google_safebrowsing}. Cryptojacking~\cite{konoth2018minesweeper,kharraz2019outguard}, fingerprinting~\cite{iqbal2021fpinspector} and scareware~\cite{rajab2010nocebo} each received their own detectors and corpora. These systems are far more mature than ours on their own threat and their own data, and \bisbench{} is not comparable with their corpora (Section~\ref{sec:limits}). What they share is that the feature set encodes the threat: a new class is a new pipeline. BIS inverts this. One general security model is asked one typed question per class, so adding a class is adding a question, and the model runs locally with no blocklist or server in the loop.

\paragraph{Artificial immune systems.} The analogy that names this work has a history. Forrest et al.\ framed intrusion detection as self/nonself discrimination~\cite{forrest1994self}, Kephart proposed a computer immune system with innate and adaptive components for virus defence~\cite{kephart1994biologically}, and Hofmeyr and Forrest gave the architecture of a distributed artificial immune system~\cite{hofmeyr2000architecture}. BIS borrows the two-tier structure, a fast inexpensive innate layer that runs always and a slow specific adaptive layer that runs when triggered, and realises both tiers with the two inference regimes of one transformer rather than with detector populations.

\paragraph{Reading answers from inside the model.} The \sone{} readout we deploy reads a decision from the hidden state at marker positions through a bilinear pointer head~\cite{vinyals2015pointer,typesafe2026systemone,mayoralvilches2026transformer} instead of generating it. Related readouts fine-tune a model for classification from its internal state~\cite{li2023labelsupervised}, compare prompted answers with probability measurements~\cite{hu2023prompting}, and show that probes on hidden states know more than the generated text reveals~\cite{orgad2024llmsknow}; distilling deliberate reasoning into direct answers is a parallel line~\cite{yu2024distilling}. Recent work on the same typed-decision lineage studies calibration and selective automation for agent security decisions~\cite{liu2026systemonesecurity,wu2026reflex}. We add the deployment regime those works assume away: the readout in a browser, on quantised weights, with the exact normalisation and option softmax it was trained with.

\paragraph{Models in the browser, and what compression does to answers.} Several runtimes execute language models on WebGPU: ONNX Runtime Web and GenAI~\cite{onnxruntime,onnxruntimegenai}, Transformers.js~\cite{transformersjs}, WebLLM~\cite{ruan2024webllm} and newer multi-precision engines~\cite{levine2026llamasweb}; dispatch overhead dominates token-by-token generation on WebGPU~\cite{maczan2026webgpu}, and measurement studies characterise browser~\cite{wang2024anatomizing} and mobile~\cite{laskaridis2024melting} inference by speed and memory rather than by fidelity of answers. Compression can preserve aggregate accuracy while changing many individual answers~\cite{dutta2024accuracy}, mostly where the full model was unsure~\cite{proskurina2024quantization}, with effects on internal states that depend on method and width~\cite{wang2026compressedlens}. Security-specialist models are evaluated at full precision~\cite{kassianik2025foundationsec} or report quantisation losses on generative question answering~\cite{gungor2025aqua}. Worrel's study~\cite{worrel2026system1edge} is the direct predecessor of this paper: it is the first to measure a cybersecurity transformer's decisions across browser runtimes and precisions, and Section~\ref{sec:prior} summarises it.

\section{Prior Work: \sone{} at the Edge}\label{sec:prior}
This section summarises the \sone{}-at-the-edge line of work~\cite{worrel2026system1edge,worrel2026bestbuild,worrel2026runner,worrel2026builds}, whose builds, runner, data and findings this paper reproduces, corrects where its own archive records a correction, and extends. We state its results as published and note where our own measurements bear on them.

\paragraph{Question and method.} Worrel asked whether the model's \sone{} answers survive the compression needed to fit a browser. Eighteen builds of \aliasmicro{} (\texttt{alias3-micro-1026}, the public release~\cite{aliasmicro1026card}) were produced on four runtimes, ONNX Runtime GenAI, WebLLM/MLC, Transformers.js and a decode-only ONNX graph, at precisions from fp32 to 4-bit, and each was compared with the original PyTorch bf16 model on a $172$-question decision test set written for the study ($65$ yes/no, $57$ multiple choice, $50$ severity levels), with exact McNemar tests, Holm correction, bootstrap intervals and \emph{flip rates}, the share of questions whose answer differs from the original's. One library, the study's browser runner~\cite{worrel2026runner}, drove all four runtimes.

\paragraph{Findings.} Three results carry over to this paper. (i)~\emph{Accuracy survives, answers move.} None of the $180$ paired tests reached $p<0.05$ (smallest $p=0.0625$), so no build is detectably less accurate than the original at that sample size; yet the 4-bit builds change $5.8$--$13.4\%$ of answers read from the logits and $12.2$--$19.8\%$ read through the head, and dynamic W8A8 builds up to $28.5\%$. Only 4-bit builds with a quantised embedding fit the browser's per-file limit, at $2.37$--$3.04$\,GB, because the $248{,}320$-token embedding alone is $2.54$\,GB in fp32. (ii)~\emph{The official head could be recovered only in part.} The released \texttt{head.pt} was truncated ($4{,}961{,}280$ of $5{,}248{,}703$ bytes, cut inside the third tensor); by parsing the zip stream Worrel recovered $100\%$ of the query projection, $89.1\%$ of the key weights and none of the key bias, replaced the lost bias by a constant offset ($+16.27$), found the read positions by search (query at \texttt{<decide>}, key at the last position for yes/no and at each \texttt{</opt>} for choice), and, because the marker-to-token mapping was unpublished, wrote the markers as plain text in a ChatML template. All head results of that study are therefore results for a partial head on an approximate interface, as its author states. (iii)~\emph{The question format is a free lever.} On $161$ real JavaScript items labelled from CodeQL security tests and OWASP Juice Shop~\cite{github_codeql,owasp_juiceshop}, asking one yes/no question per vulnerability class and keeping the maximum raised AUROC from $0.764$ to $0.831$ for the head and from $0.807$ to $0.847$ for the logits, with no training. A separate ridge readout on the hidden state, \texttt{head\_score.bin}, fitted on $30$ synthetic cases, ranked severity with Spearman $0.849$ on held-out cases and reached $0.52$ exact and $0.94$ within one level on the $50$ test questions, against $0.40$ and $0.70$ for the logits; a second version in the research archive reaches $0.98$ within one level.

\begin{table}[t]
\centering\bistable\scriptsize\setlength{\tabcolsep}{2.6pt}
\caption{\textbf{Browser-fitting builds of \aliasmicro{} in the prior study}~\cite{worrel2026system1edge}, compared with the original model on its $172$-question decision set. Comp.\ is the \sone{} composite (mean of yes/no and choice accuracy through the recovered head and severity within $\pm1$ through \texttt{head\_score.bin}); flips are the share of questions whose answer differs from the original's (logits read / head read). Time per decision is the measured value in Chrome on an Apple M2 Pro from the study's research archive, not the printed one where the two differ (Section~\ref{sec:prior}). The Transformers.js q4f16 time is the study's median with the part of that run slowed by other jobs dropped ($70$ of $172$ questions kept; the median over the whole run is $20.05$\,s). No build differs significantly from the original; the shaded row is the build \aliasweb{} uses.}
\label{tab:prior}
\begin{tabular}{@{}llrrrr@{}}
\toprule
\tabhead \textbf{Runtime} & \textbf{Build} & \textbf{GB} & \textbf{Comp.} & \textbf{Flips l/h (\%)} & \textbf{s/dec.} \\
\midrule
reference & PyTorch bf16 & -- & 0.859 & -- & -- \\
\tabalt decode-only & q4f16 & 2.73 & 0.879 & 12.8 / 12.2 & 21.2 \\
WebLLM & q4f32\_1 & 2.37 & 0.873 & 13.4 / 14.0 & 5.78 \\
\rowcolor{fig_s1!14} GenAI & int4, b32 & 2.44 & 0.871 & 11.6 / 14.5 & 0.54 \\
decode-only & q4 & 3.03 & 0.871 & 10.5 / 14.5 & -- \\
\tabalt Transformers.js & q4 & 3.04 & 0.871 & 10.5 / 14.5 & 7.43 \\
Transformers.js & q4f16 & 2.73 & 0.866 & 9.9 / 14.5 & 5.75 \\
\tabalt decode-only & bnb4-nf4 & 2.73 & 0.864 & 6.4 / 12.2 & -- \\
WebLLM & q4f16\_1 & 2.37 & 0.858 & 10.5 / 19.8 & 4.25 \\
\tabalt Transformers.js & bnb4 & 2.74 & 0.841 & 5.8 / 12.8 & -- \\
\bottomrule
\end{tabular}
\end{table}

\paragraph{Recommended build and recipe.} The accompanying brief~\cite{worrel2026bestbuild} recommends the ONNX Runtime GenAI int4 (block $32$) build for WebGPU: $2.44$\,GB, the same accuracy as the other browser-fitting builds, official tooling, and one prefill per decision, against token-by-token execution for the decode-only graph and one prefill per read position for WebLLM. Its recipe (\texttt{onnxruntime-genai} $0.18$ with \texttt{include\_hidden\_states}) is the one we execute in Section~\ref{sec:model-build}, with the caveat the brief itself records: on Apple Silicon the int4 rounding and the rotary tables differ from the published files in about $0.02\%$ of the int4 bytes ($5$--$8\%$ of the rotary-table bytes), and the build must be run on Linux x86\_64 for bit-exact weights.

\paragraph{Correction recorded by the research archive.} The sources distributed with the study document one timing value in its Figure~3 that its measurement files do not support: the decode-only q4f16 build is printed at $9$--$14$\,s per decision, which matches recorded model-load times, where the measured value is $21.2$\,s per question. The fastest build is unaffected: GenAI int4 at $0.54$\,s median per question in Chrome on an Apple M2 Pro ($0.67$\,s when the whole run is divided by its $202$ prompts). Table~\ref{tab:prior} and Figure~\ref{fig:latency}a use the measured values. Two methodological points also matter for what follows: the study's accuracy figures come from server runs of the same files (choice $84.2\%$ for GenAI int4 on CUDA against $80.7\%$ in Chrome, with both runs in the archive), and the per-decision time is defined differently in the paper and in the brief, so cross-runtime speed comparisons mix the effect of the runtime with that of the graph.

What the prior work could not do, for want of the official pieces, is the subject of the next section: run \sone{} in the browser exactly as it was trained.

\section{\aliasweb{}: a Cybersecurity Transformer for the Browser}\label{sec:model}
\aliasweb{} is a cybersecurity transformer prepared for the browser: a specific checkpoint of the \aliasmicro{} lineage, the exact readout it was trained with, and a 4-bit graph that runs both inference regimes on WebGPU. This section gives the derivation (\S\ref{sec:model-lineage}), the official pieces that the prior work lacked (\S\ref{sec:model-pieces}), the build (\S\ref{sec:model-build}), the two regimes as they run in a tab (\S\ref{sec:model-dual}), and the validation and cost measurements (\S\ref{sec:model-validation}, \S\ref{sec:model-privacy}).

\subsection{Lineage}\label{sec:model-lineage}
The backbone is that of \aliasmicro{}~\cite{mayoralvilches2026transformer}: a hybrid \basemodel{}-derived transformer of $32$ layers at hidden width $2560$ ($24$ Gated DeltaNet mixers and $8$ Gated-Attention layers with $4$ key-value heads) and a tied $248{,}320$-token embedding, specialised for security by two-phase on-policy distillation~\cite{agarwal2024onpolicy} on the CAI dataset~\cite{mayoralvilches2026caidataset}. Two low-rank adapters~\cite{hu2022lora} are fused into it. The first is the parent model's \emph{decision adapter}, a rank-$16$ LoRA trained together with the bilinear pointer head that implements \sone{}. The second, which distinguishes \aliasweb{} from the public release, is a \emph{context-category sensor}: a rank-$16$ adapter trained for two epochs at learning rate $5\times10^{-5}$ on $12{,}576$ replayed decision examples and $4{,}172$ context-category examples drawn from capture-the-flag campaigns, with the head frozen (\texttt{head\_lr}$=0$) so that the decision head is inherited unchanged. The sensor teaches the backbone to tell in which operational context a piece of evidence sits. BIS does not use that ability directly, since it asks yes/no questions only, and we did not compare this checkpoint with the public release on its battery. The fused bf16 weight set is $8.4$\,GB; the complete head file is $5{,}248{,}703$ bytes, the size the truncated public file should have had (Section~\ref{sec:prior}). We call the resulting artefact \aliasweb{} (release \texttt{alias-web-1126}, published as \texttt{alias3-web} with its readout files, library and demo page at \href{\aliasweburl}{\texttt{huggingface.co/\allowbreak aliasrobotics/\allowbreak alias3-web}}); throughout, ``the model'' means this checkpoint.

\subsection{The official pieces of \sone{}}\label{sec:model-pieces}
Three pieces make the browser run \sone{} exactly as it was trained. Each was missing from, or approximated in, the prior deployment, and Table~\ref{tab:vsedge} sets the two side by side.

\paragraph{The marker-to-token mapping.} The typed interface frames a decision as one causal sequence, \texttt{<state>}\,$s$\,\texttt{<q>}\,$q$\,\texttt{<opt>}\,$o_1$\,\texttt{</opt>}\allowbreak\ldots\allowbreak\texttt{<opt>}\,$o_K$\,\texttt{</opt>}\,\allowbreak\texttt{<decide>}. The markers are not new vocabulary: they reuse five rarely used special tokens of the base tokenizer, listed in Table~\ref{tab:markers}, and the decision adapter gives them their meaning. Because the mapping was unpublished, the prior work wrote the markers as plain text inside a chat template, so that the head read positions whose embeddings it had never seen in training. \aliasweb{} encodes the ids directly, and sanitises user-supplied text by rewriting any literal \texttt{<|name|>} to \texttt{<\textbrokenbar name\textbrokenbar>}, the same rule as in training, so that no page content can forge an option boundary. The single causal row is mathematically equivalent to the branch-masked packing used in training, which is why the browser needs no custom attention masks.

\begin{table}[t]
\centering\bistable\footnotesize\setlength{\tabcolsep}{4pt}
\caption{\textbf{The official \sone{} markers}: existing special tokens of the base tokenizer, reused without adding embedding rows. The head reads the final hidden state at \texttt{<decide>} (query) and at each \texttt{</opt>} (key).}
\label{tab:markers}
\begin{tabular}{@{}llrl@{}}
\toprule
\tabhead \textbf{Marker} & \textbf{Token} & \textbf{Id} & \textbf{Role in the row} \\
\midrule
\texttt{<state>} & \texttt{<|fim\_prefix|>} & 248060 & opens the evidence \\
\tabalt \texttt{<q>} & \texttt{<|fim\_middle|>} & 248061 & opens the question \\
\texttt{<opt>} & \texttt{<|box\_start|>} & 248049 & opens an option \\
\tabalt \texttt{</opt>} & \texttt{<|box\_end|>} & 248050 & closes an option: key read here \\
\texttt{<decide>} & \texttt{<|fim\_suffix|>} & 248062 & closes the row: query read here \\
\bottomrule
\end{tabular}
\end{table}

\paragraph{The complete head, and the normalisation it expects.} Writing $h_x\in\mathbb{R}^{2560}$ for the hidden state at token $x$ \emph{after} the final RMSNorm, the pointer head~\cite{vinyals2015pointer} scores option $i$ as
\begin{equation}
z_i=\frac{q(h_{\texttt{<decide>}})^{\!\top}k(h_{\texttt{</opt>}_i})}{\sqrt{d}\,T},\qquad p=\operatorname{softmax}(z),
\label{eq:head}
\end{equation}
with affine projections $q,k:\mathbb{R}^{2560}\to\mathbb{R}^{256}$ ($1.3$\,M parameters, complete in \aliasweb{}: weights and biases of both projections) and the softmax taken over the $K$ options for every question type, a yes/no question being the $K=2$ case; the prior deployment applied a sigmoid to a single score at the last position, which is not the trained form. $T$ is the head temperature stored in the checkpoint ($1.0$ for \aliasweb{}, uncalibrated; the argmax is invariant to it). The subtlety that cost the most during bring-up is the normalisation. The exported graph returns \texttt{hidden\_states} \emph{before} the final RMSNorm, while the head was trained on the normalised state. Fed pre-norm states, the head sees magnitudes about five times smaller than in training, its scores collapse to the order of $\mathrm{e}^{-52}$ and the argmax is noise (Appendix~\ref{app:engineering}, B3). \aliasweb{} exports the $2560$ norm weights $w$ as a $10$\,KB file (\texttt{norm.bin}) and applies
\begin{equation}
\tilde h=\frac{h}{\sqrt{\tfrac{1}{2560}\sum_j h_j^2}}\odot w
\label{eq:norm}
\end{equation}
in JavaScript before Eq.~\eqref{eq:head}. With it, the 4-bit graph in the browser answers the validation battery as expected and returns the decision direction of the bf16 model in PyTorch on the prompt we compared (\S\ref{sec:model-validation}). The prior work did not need this step because its build exposed the post-norm node; ours exposes the pre-norm one, and the lesson generalises: an exported hidden state must be checked for the normalisation the head expects.

\begin{table*}[t]
\centering\bistable\footnotesize
\caption{\textbf{What \aliasweb{} changes with respect to the prior edge deployment, and why.} The runner is the prior work's, vendored unchanged; the readout layer is new.}
\label{tab:vsedge}
\begin{tabularx}{\textwidth}{@{}l Y Y Y@{}}
\toprule
\tabhead \textbf{Piece} & \textbf{prior deployment~\cite{worrel2026system1edge}} & \textbf{\aliasweb{} (this work)} & \textbf{Why} \\
\midrule
\sone{} markers & plain text \texttt{<state>}\ldots\ inside a chat template & the five official token ids (Table~\ref{tab:markers}) & the mapping was unpublished; the head had never seen plain-text markers in training \\
\tabalt Head & recovered from the truncated \texttt{head.pt}: $100\%$ of $q$, $89.1\%$ of $k$'s weights, $k$'s bias replaced by a constant offset ($+16.27$) & complete head from the training run, fp16 & the public file is cut inside the third tensor; the complete one exists \\
Yes/no readout & $q$ at \texttt{<decide>} $\cdot$ $k$ at the last position, sigmoid & softmax over the two official options, $k$ at each \texttt{</opt>} & the trained form: every question is a choice \\
\tabalt Severity & \texttt{head\_score.bin}, a separate ridge readout (proof of concept) & official $1$--$5$ choice through the same head & one mechanism; calibration remains future work (\S\ref{sec:model-validation}) \\
Final norm & implicit: that build exposed the post-norm node & explicit \texttt{norm.bin} applied in JavaScript (Eq.~\ref{eq:norm}) & this build exports pre-norm states; without the norm the head saturates (B3) \\
\tabalt Runner & the study's browser runner (MIT) & the same, vendored & tested across $18$ builds; our own port produced NaNs (B2) \\
Model & \aliasmicro{} (\texttt{1026}, public release) & \aliasweb{}: \texttt{1026} + decision adapter + context-category sensor, fused & the latest checkpoint of the lineage, with its complete head; not compared with the public release on the battery \\
\bottomrule
\end{tabularx}
\end{table*}

\subsection{Build}\label{sec:model-build}
The graph follows the recipe of the prior brief~\cite{worrel2026bestbuild}: ONNX Runtime GenAI's model builder (\texttt{onnxruntime-genai} $0.18$) with \texttt{-p int4 -e webgpu} and \texttt{include\_hidden\_states=true}, run on Linux x86\_64 so that the int4 rounding and rotary tables match the published \aliasmicro{} builds byte for byte (on Apple Silicon about $0.02\%$ of the int4 bytes and $5$--$8\%$ of the rotary-table bytes differ, with cosine $\approx0.98$ at the positions the head reads and the same accuracy, as the re-run of the recipe in the prior study's research archive records). The build takes $16$\,s and produces \texttt{model.onnx} with $2{,}440{,}274{,}944$ bytes of external data: $2.44$\,GB, identical in size to the published \aliasmicro{} int4 build. Two small files complete it, \texttt{head.bin} ($2{,}622{,}545$ bytes, the complete head in fp16 with a JSON header in the runner's format) and \texttt{norm.bin}. The runtime is the prior study's browser runner~\cite{worrel2026runner} (MIT), vendored unchanged: it chunks the prompt in blocks of $128$ tokens, feeds and drains the graph's $64$ cache inputs ($8$ attention layers with key/value caches of head dimension $256$ and $24$ DeltaNet layers with convolution states of shape $[1,8192,3]$ and recurrent states), and exposes the hidden states at requested positions; a port of our own produced NaNs in $398$ of $2560$ hidden dimensions because it mapped the dynamic DeltaNet cache dimensions to the attention head dimension, and we kept the battle-tested runner (B2). The int4 graph runs only on the WebGPU execution provider, since the \texttt{GatherBlockQuantized} kernel that gathers the quantised embedding has no CPU implementation, which also means the graph cannot be debugged on CPU (B5). On top of the runner, a $10$\,KB library (\texttt{aliasweb.mjs}) implements the official encoding, the normalisation, the head and the three question kinds: \texttt{noul} (yes/no), \texttt{choice} ($K$ options) and \texttt{score} (an ordinal $1$--$5$ scale as a $K=5$ choice), returning the option probabilities, the raw scores and the latency.

\subsection{Two regimes from one loaded graph}\label{sec:model-dual}
\begin{figure*}[t]
\centering
\resizebox{\textwidth}{!}{%
\figfonts
\begin{tikzpicture}[x=1cm,y=1cm]
\node[card=fig_control, minimum width=2.75cm, minimum height=1.6cm, anchor=west] (inp) at (0,3.3) {};
\draw[plain] (0.2,2.75) rectangle (0.95,3.85); \draw[fig_rule, line width=.8pt] (0.32,3.62)--(0.83,3.62) (0.32,3.42)--(0.83,3.42) (0.32,3.22)--(0.72,3.22) (0.32,3.02)--(0.83,3.02);
\node[tt, anchor=west] at (1.08,3.7) {state $s$};
\node[tx, anchor=north west, text width=1.62cm, align=left] at (1.08,3.52) {a page observation, code or a log; it never leaves the machine};
\node[chip, minimum width=10.5cm, minimum height=1.35cm, anchor=west] (bb) at (3.2,3.3) {};
\foreach \x in {3.7,4.3,...,13.4} {\node[pin, minimum width=1.4pt, minimum height=2.6pt] at (\x,4.0) {}; \node[pin, minimum width=1.4pt, minimum height=2.6pt] at (\x,2.6) {};}
\node[tt] at (8.45,3.62) {shared backbone $\cdot$ \texttt{alias-web} int4 (2.44 GB) on WebGPU $\cdot$ one weight set, loaded once};
\node[ts] at (8.45,3.3) {32 layers: 24 Gated DeltaNet + 8 Gated Attention $\cdot$ hidden 2560 $\cdot$ vocabulary 248\,320 $\cdot$ ORT-WebGPU 1.30};
\node[ts] at (8.45,3.0) {each prefill returns \texttt{hidden\_states} (pre-norm, at the requested positions) and logits (last position)};
\draw[bus] (2.75,3.3) -- (3.2,3.3);
\draw[rounded corners=4pt, fill=fig_s1!7, draw=fig_s1, line width=.7pt] (3.15,4.5) rectangle (13.8,6.95);
\node[tt, anchor=west, text=fig_s1!70!black] at (3.3,6.75) {System 1 $\cdot$ one typed question, one prefill, zero tokens decoded};
\node[card=fig_s1, minimum width=2.75cm, minimum height=1.35cm, anchor=west] (s1a) at (3.35,5.725) {};
\node[tt, anchor=west] at (3.45,6.18) {encode};
\node[tm, font=\ttfamily\fontsize{4.7}{5.7}\selectfont, text=fig_ink, anchor=north west, text width=2.6cm, align=left] at (3.45,5.98) {<state>\,$s$\,<q>\,$q$\\<opt>Yes</opt><opt>No</opt>\\<decide>};
\node[tx, anchor=north west] at (3.45,5.35) {official ids $\cdot$ user text sanitised};
\node[card=fig_s1, minimum width=1.55cm, minimum height=1.35cm, anchor=west] (s1b) at (6.3,5.725) {};
\node[tt, anchor=west] at (6.4,6.18) {prefill $\times1$};
\node[tx, anchor=north west, text width=1.38cm, align=left] at (6.4,5.98) {hidden states at \texttt{<decide>} and at each \texttt{</opt>}};
\node[card=fig_thinking, minimum width=1.55cm, minimum height=1.35cm, anchor=west] (s1c) at (8.05,5.725) {};
\node[tt, anchor=west] at (8.15,6.18) {norm.bin};
\node[tx, anchor=north west, text width=1.38cm, align=left] at (8.15,5.98) {final RMSNorm (Eq.~\ref{eq:norm}): the state the head saw in training};
\node[card=fig_thinking, minimum width=2.45cm, minimum height=1.35cm, anchor=west] (s1d) at (9.8,5.725) {};
\node[tt, anchor=west] at (9.9,6.18) {head.bin};
\node[font=\fontsize{5.4}{6}\selectfont, text=fig_ink, anchor=west] at (9.9,5.78) {$z_i \propto q(\tilde h_{\mathrm{dec}})^{\!\top} k(\tilde h_i)$};
\node[tx, anchor=north west, text width=2.3cm, align=left] at (9.9,5.56) {the complete bilinear head, scaled as in Eq.~\ref{eq:head}};
\node[card=fig_s1, minimum width=1.2cm, minimum height=1.35cm, anchor=west] (s1e) at (12.45,5.725) {};
\node[tt, anchor=west] at (12.55,6.18) {softmax};
\node[tx, anchor=north west, text width=1.05cm, align=left] at (12.55,5.98) {$P(\mathrm{yes})$\\0.822 on the XSS};
\foreach \a/\b in {s1a/s1b, s1b/s1c, s1c/s1d, s1d/s1e} \draw[flow1] (\a.east) -- (\b.west);
\draw[flow1] (7.1,4.0) -- (7.1,5.05);
\node[lab, anchor=west] at (7.15,4.45) {one forward pass};
\node[ts, anchor=east] at (13.7,4.72) {$\approx$152 ms per question $\cdot$ head 1.18 ms $\cdot$ the cost is the prefill};
\draw[flow1] (2.5,3.9) to[out=90,in=180] (3.35,5.725);
\draw[rounded corners=4pt, fill=fig_s2!7, draw=fig_s2, dashed, line width=.7pt] (3.15,-0.55) rectangle (13.8,2.1);
\node[tt, anchor=west, text=fig_s2] at (3.3,1.9) {System 2 $\cdot$ autoregressive generation from the same weights, on request};
\node[card=fig_s2, minimum width=2.75cm, minimum height=1.3cm, anchor=west] (s2a) at (3.35,0.9) {};
\node[tt, anchor=west] at (3.45,1.33) {chat prompt};
\node[tm, font=\ttfamily\fontsize{4.7}{5.7}\selectfont, text=fig_ink, anchor=north west, text width=2.6cm, align=left] at (3.45,1.13) {<|im\_start|>user\,\ldots<|im\_end|>\\assistant</think>};
\node[tx, anchor=north west, text width=2.55cm, align=left] at (3.45,0.68) {the closing marker is forced: no open deliberation};
\node[card=fig_s2, minimum width=1.55cm, minimum height=1.3cm, anchor=west] (s2b) at (6.3,0.9) {};
\node[tt, anchor=west] at (6.4,1.33) {prefill};
\node[tx, anchor=north west, text width=1.38cm, align=left] at (6.4,1.13) {logits over 248\,320 tokens at the last position};
\node[card=fig_s2, minimum width=1.55cm, minimum height=1.3cm, anchor=west] (s2c) at (8.05,0.9) {};
\node[tt, anchor=west] at (8.15,1.33) {sample};
\node[tx, anchor=north west, text width=1.38cm, align=left] at (8.15,1.13) {temperature 0.7\\top-$p$ 0.9\\append, repeat};
\draw[flow2] (8.8,0.0) to[out=-90,in=-90,looseness=1.4] (7.1,0.0);
\node[lab, text=fig_s2] at (7.95,-0.3) {one forward pass per token};
\node[card=fig_s2, minimum width=3.85cm, minimum height=1.3cm, anchor=west] (s2d) at (9.8,0.9) {};
\node[tt, anchor=west] at (9.9,1.33) {text};
\node[font=\sffamily\itshape\fontsize{5.3}{6.3}\selectfont, text=fig_ink, anchor=north west, text width=3.65cm, align=left] at (9.9,1.13) {``Cross-Site Scripting\ldots\ sanitize using DOMPurify, or set it as a data attribute\ldots''};
\foreach \a/\b in {s2a/s2b, s2b/s2c, s2c/s2d} \draw[flow2] (\a.east) -- (\b.west);
\draw[flow2] (7.1,2.6) -- (7.1,1.55);
\node[ts, anchor=east] at (13.7,2.3) {$\approx$96 ms per token (GPU dispatch bound) $\cdot$ 10.5 tok/s $\cdot$ 11--73 tokens};
\draw[flow2] (2.5,2.7) to[out=-90,in=180] (3.35,0.9);
\node[tt, anchor=west] at (14.05,6.78) {cost per answer};
\node[tx, anchor=west] at (14.05,6.54) {measured on the same three snippets};
\node[ts, anchor=west] at (14.05,6.25) {System 1 decision};
\fill[fig_s1, rounded corners=1pt] (14.05,5.85) rectangle (14.17,6.08); \node[ts, anchor=west] at (14.25,5.965) {0.15--0.20 s};
\node[ts, anchor=west] at (14.05,5.45) {System 2 answer};
\fill[fig_s2, rounded corners=1pt] (14.05,5.05) rectangle (16.8,5.28); \node[ts, anchor=west, text=white] at (14.15,5.165) {1.1--6.9 s};
\node[font=\sffamily\bfseries\fontsize{8}{9}\selectfont, text=fig_ink, anchor=west] at (14.05,4.7) {6--39$\times$};
\node[ts, anchor=west, text width=3.05cm] at (14.05,3.95) {a typed decision fits in a scroll or a page load; a generated answer does not. BIS runs System 1 on every page and System 2 only when asked.};
\node[ts, anchor=west, text width=3.05cm] at (14.05,2.75) {both regimes read the same loaded graph: System 1 the hidden states at the markers, System 2 the logits, fed back token by token.};
\draw[fig_s1, line width=.9pt] (0.1,-0.9) -- (0.6,-0.9); \node[ts, anchor=west] at (0.68,-0.9) {System 1 flow (one prefill)};
\draw[fig_s2, line width=.9pt, dashed] (3.3,-0.9) -- (3.8,-0.9); \node[ts, anchor=west] at (3.88,-0.9) {System 2 flow (one pass per token)};
\node[ts, anchor=west] at (7.7,-0.9) {\textcolor{fig_thinking}{\rule{.6em}{.6em}} official readout pieces (this work)};
\end{tikzpicture}}
\caption{\textbf{The two inference regimes of \aliasweb{} as they run in the browser}, around one loaded 4-bit weight set (centre). \emph{Top}, \sone{}: the evidence, the question and the options are encoded with the official marker tokens in one causal row; one prefill returns the hidden states at \texttt{<decide>} and at each \texttt{</opt>}; the final RMSNorm (\texttt{norm.bin}) and the complete bilinear head (\texttt{head.bin}), the two official readout pieces in ochre, score the options and a softmax returns one probability per option, in about $152$\,ms with zero tokens decoded. \emph{Bottom}, \stwo{}: a chat prompt with a forced end-of-deliberation prefix is prefilled, and the logits over the $248{,}320$-token vocabulary are sampled token by token, about $96$\,ms per token because GPU dispatch dominates~\cite{maczan2026webgpu}. \emph{Right}: the cost measured on the same three snippets, $0.15$--$0.20$\,s for a decision against $1.1$--$6.9$\,s for an answer.}
\label{fig:dual}
\end{figure*}
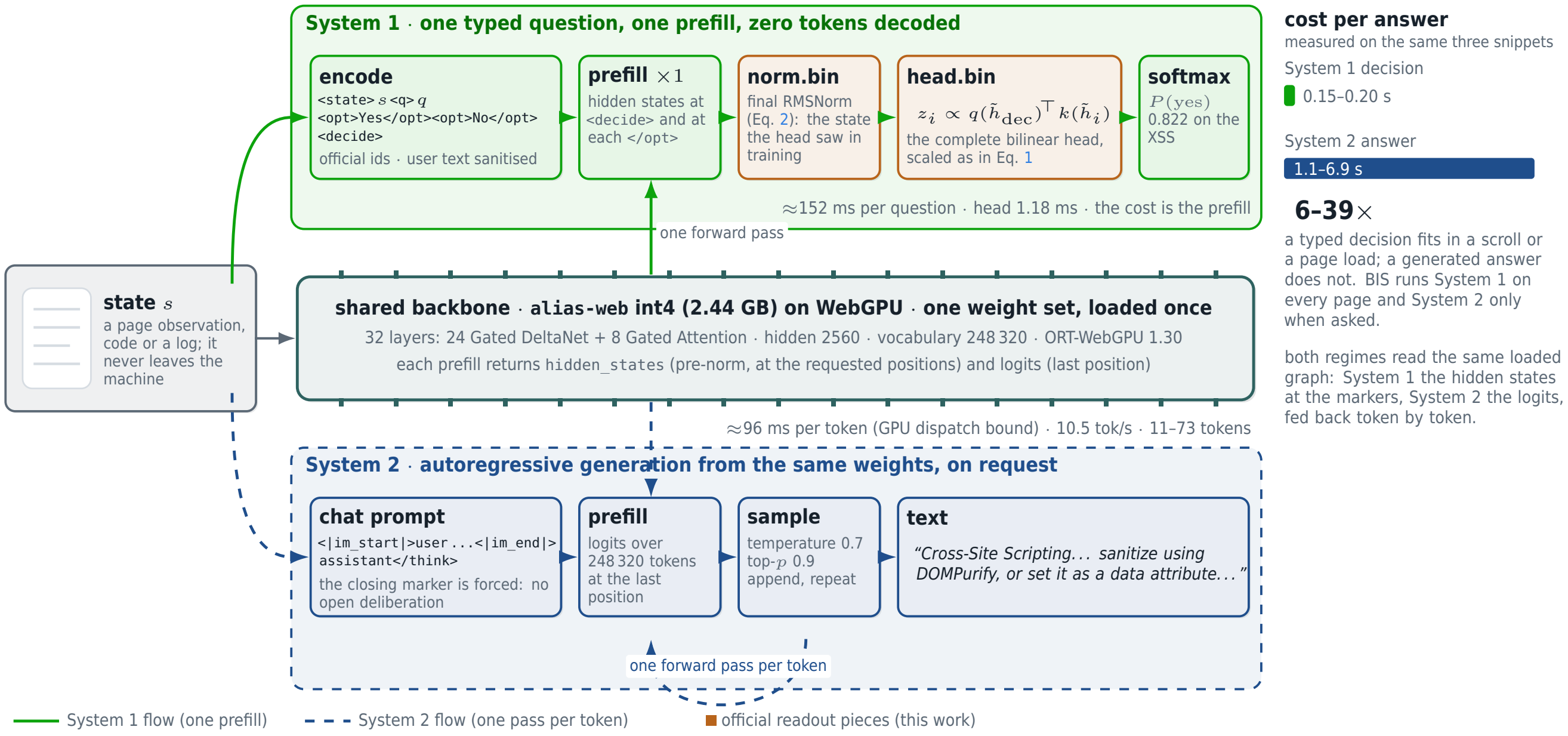
Figure~\ref{fig:dual} shows how the same loaded graph serves \sone{} and \stwo{}. \sone{} is one prefill of the marker row followed by Eqs.~\eqref{eq:norm} and~\eqref{eq:head}. \stwo{} is the generative path over the same weights: the library prefills a chat prompt and samples from the logits (temperature $0.7$, top-$p$ $0.9$) until an end token or a budget. One adaptation was necessary. The base model family deliberates by default inside a thinking block, and this checkpoint in 4-bit does not close the block: left to itself it produced $400$ tokens of deliberation without an answer, and the usual instruction to skip thinking was ignored (B8). Forcing the assistant turn to begin with the closed deliberation marker, \texttt{<|im\_start|>assistant\textbackslash n</think>\textbackslash n\textbackslash n}, yields direct, correct answers; the special tokens must be inserted by id, because the sanitiser that protects the \sone{} markers rewrites them if they pass through the tokenizer as text (B9).

Table~\ref{tab:dual} gives both regimes on three snippets in one tab, with the model already loaded. \sone{} answers in $148$--$197$\,ms; \stwo{} produces an accurate diagnosis and fix in $1.1$ to $6.9$\,s ($11$ to $73$ tokens at $10.5$ tokens/s). The ratio on the same snippet, $6$ to $39\times$, is the reason the extension of Section~\ref{sec:bis} runs \sone{} on every page and \stwo{} only on request: a generative pass does not fit inside a scroll or a page load, a typed decision does.

\begin{table}[t]
\centering\bistable\footnotesize
\caption{\textbf{\sone{} and \stwo{} on the same three snippets}, same tab, same loaded weights. \sone{} uses the generic question ``Does this code contain a security vulnerability?''; the directed question raises the XSS case to $0.822$ (Figure~\ref{fig:question}). \stwo{} is prompted as a reviewer with the forced end-of-deliberation prefix.}
\label{tab:dual}
\begin{tabularx}{\linewidth}{@{}l r r Y r r@{}}
\toprule
\tabhead \textbf{Snippet} & \textbf{S1 $P$} & \textbf{\lat{ms}} & \textbf{\stwo{} answer (generated)} & \textbf{tok} & \textbf{\lat{s}} \\
\midrule
DOM-XSS & 0.040 & 179 & ``Cross-Site Scripting (XSS) because it directly inserts user-controlled input\ldots\ sanitize using DOMPurify, or set it as a data attribute\ldots'' & 73 & 6.9 \\
\tabalt SQLi & 1.000 & 148 & ``SQL injection due to string concatenation in the query. Use parameterized queries or prepared statements\ldots'' & 30 & 2.9 \\
safe & 0.000 & 197 & ``The code is safe and does not contain any vulnerabilities.'' & 11 & 1.1 \\
\bottomrule
\end{tabularx}
\end{table}

\subsection{Validation and cost}\label{sec:model-validation}
\paragraph{Decision quality and parity with the reference.} Table~\ref{tab:battery} runs six snippets through \sone{} with one directed question each. The five vulnerable snippets score $0.822$--$1.000$ except path traversal ($0.428$), and the safe control scores $0.000$. For parity we ran one $16$-token prompt of the battery through the bf16 weights in PyTorch on a server GPU: the reference returns logits of $\pm2.4$ with the same yes/no direction as the browser's $+0.9/-3.9$ after normalisation. On that prompt the decision survives the compression, as the prior study found at scale with its flip-rate analysis (its GenAI int4 build changed $11.6\%$ of logits answers and $14.5\%$ of head answers without a detectable accuracy change); what this adds is that the agreement is observed for the \emph{official} readout on the \emph{official} interface. One prompt verifies that the readout is wired as trained. It does not measure how often the two precisions agree, which remains to be done for the official head.

\begin{table}[t]
\centering\bistable\footnotesize
\caption{\textbf{Validation battery} in Chrome (WebGPU, Apple M3 Max): one directed yes/no question per snippet, with the severity level returned by the \texttt{score} kind (its expectation over the five levels in parentheses) and the latency of the yes/no question. The first question pays the GPU just-in-time compilation.}
\label{tab:battery}
\begin{tabularx}{\linewidth}{@{}Y Y r l r@{}}
\toprule
\tabhead \textbf{Snippet} & \textbf{Question} & \textbf{$P(\text{yes})$} & \textbf{level} & \textbf{\lat{ms}} \\
\midrule
DOM-XSS\newline\texttt{innerHTML} + query parameter & Is user input written into the DOM without escaping? & \textbf{0.822} & 1 (1.44) & 202 \\
\tabalt SQLi\newline string concatenation in \texttt{db.query} & Does this code concatenate user input into a SQL query? & \textbf{1.000} & 1 (2.61) & 140 \\
Command injection\newline\texttt{exec("curl "+url)} & Does this code pass user input to a shell command? & \textbf{1.000} & 2 (2.69) & 146 \\
\tabalt Path traversal\newline\texttt{sendFile(}\allowbreak\texttt{"/var/data/"+f)} & Can user input escape the intended directory in this file path? & 0.428 & 5 (3.10) & 140 \\
Hard-coded secret\newline API key in source & Does this code contain a hardcoded secret? & \textbf{1.000} & 1 (1.46) & 145 \\
\tabalt Safe\newline pure arithmetic & Is user input written into the DOM without escaping? & \textbf{0.000} & 1 (1.10) & 138 \\
\bottomrule
\end{tabularx}
\end{table}

\begin{figure}[t]
\centering
\includegraphics[width=\linewidth]{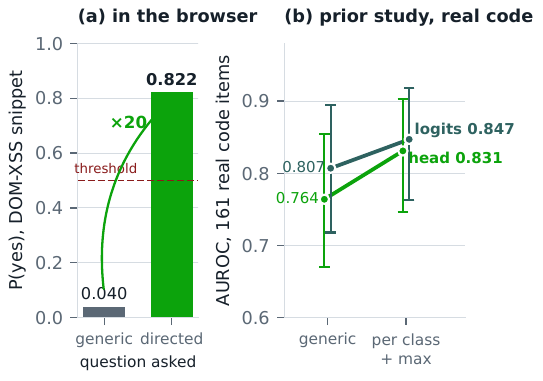}
\caption{\textbf{Asking per class is the lever}, at two scales. (a) In the browser, the same DOM-XSS snippet scores $0.040$ under the generic question and $0.822$ under the directed one (Table~\ref{tab:battery}); the dashed line is the decision threshold. (b) In the prior study, on $161$ real code items, one question per class with the maximum taken raises AUROC for both readouts~\cite{worrel2026system1edge}; whiskers are $95\%$ cluster-bootstrap intervals.}
\label{fig:question}
\end{figure}

\paragraph{The question is the lever.} Figure~\ref{fig:question} isolates the effect that the prior study found on real code and that BIS turns into a design rule. On the XSS snippet the generic question ``Does this code contain a security vulnerability?'' returns $P=0.040$; the directed question of Table~\ref{tab:battery} returns $0.822$. The same head, the same state, a different question: the readout is conditioned on what is asked. On $161$ real JavaScript items the prior study measured the same lever as AUROC $0.764\to0.831$ for the head and $0.807\to0.847$ for the logits when one question per class replaces one generic question~\cite{worrel2026system1edge}. \sone{} is therefore not a vulnerability detector with a fixed decision boundary; it is a decision procedure whose accuracy depends on the specificity of the question, and a system built on it should ask per class.

\paragraph{What the typed readouts of this checkpoint mean.} Two cautions follow from the lineage. The \texttt{choice} kind answers with the categories the sensor was trained on, the context categories of capture-the-flag campaigns, not a vulnerability taxonomy: asked to pick among weakness classes it returns \texttt{sql-injection} with probability $0.9988$ for any injectable snippet, a saturated and reproducible but semantically misdirected answer. And the \texttt{score} kind orders severity only weakly (Table~\ref{tab:battery}: path traversal at an expectation of $3.10$, above command injection at $2.69$) because the inherited head was never calibrated for severity, exactly as the prior study found for the recovered head ($5/7$ at best). The mechanism, a softmax over official options, is correct and discriminative; the semantics are those of the trained head. Calibrated severity on this graph is a matter of running the prior work's \texttt{head\_score} recipe on it (about $40$ CPU-minutes), which we leave as future work; BIS uses only the \texttt{noul} kind, where the checkpoint is strong.

\begin{figure*}[t]
\centering
\includegraphics[width=\textwidth]{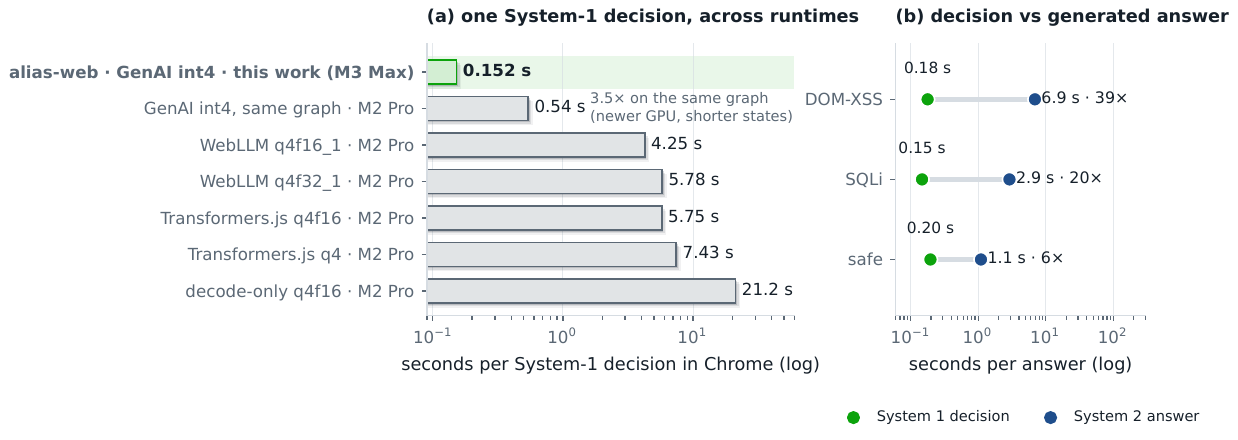}
\caption{\textbf{Cost of a decision in the browser.} (a) Time per \sone{} decision across browser runtimes: \aliasweb{} on GenAI int4 (this work, Chrome on an Apple M3 Max, $16$-token states) against the prior study's measured values in Chrome on an Apple M2 Pro~\cite{worrel2026system1edge}, in grey; the same graph runs $3.5\times$ faster here, a difference of hardware and prompt length, so the comparison is indicative. (b) \sone{} against \stwo{} on the three snippets of Table~\ref{tab:dual}, logarithmic scale: a typed decision costs $0.15$--$0.20$\,s, a generated answer $1.1$--$6.9$\,s, $6$ to $39\times$ more.}
\label{fig:latency}
\end{figure*}

\paragraph{Latency and load.} Table~\ref{tab:load} and Figure~\ref{fig:latency} give the cost. With the weights on a local server, a repeat load has the whole model ready in $2.33$\,s (runtime import $\approx0.3$\,s; reading the weights, creating the session and uploading them to the GPU $\approx1.8$\,s; heads and warm-up $\approx0.2$\,s); a first load in a fresh browser profile, which also fetches the runtime modules and compiles the GPU pipelines, took $2.5$ to $4.0$\,s in the checks of the published files. The server marks its responses as not cacheable, so every load reads the $2.44$\,GB again, which takes $0.50$\,s by itself on the local machine; over a network the same read would take about $20$\,s at $1$\,Gb/s and more than three minutes at $100$\,Mb/s, and we did not measure a remote load. A \sone{} question costs $110$--$200$\,ms on $16$-token states ($152$\,ms mean over the twelve questions of the validation session, a yes/no and a severity question per snippet of Table~\ref{tab:battery}, $1.83$\,s in all), of which the head itself is $1.18$\,ms: the cost is the prefill, not the readout. Against the prior study's measured per-decision times in Chrome on an Apple M2 Pro (GenAI int4 $0.54$\,s, WebLLM $4.3$--$5.8$\,s, Transformers.js $5.8$--$7.4$\,s, decode-only $21$\,s), the same GenAI int4 graph on an M3 Max is about $3.5\times$ faster on shorter prompts (Figure~\ref{fig:latency}a); the comparison mixes hardware and prompt length and is indicative only. \stwo{} runs at $10.5$ tokens/s on the three snippets of Table~\ref{tab:dual} ($114$ tokens in $10.9$\,s), about $96$\,ms per token, the regime where WebGPU dispatch overhead dominates~\cite{maczan2026webgpu}.

\begin{table}[t]
\centering\bistable\footnotesize
\caption{\textbf{Load and latency} of \aliasweb{} in Chrome (WebGPU, Apple M3 Max), weights served from a local Range-capable server.}
\label{tab:load}
\begin{tabularx}{\linewidth}{@{}Y r Y@{}}
\toprule
\tabhead \textbf{Phase} & \textbf{\lat{time}} & \textbf{Conditions} \\
\midrule
Reading the weights ($2.44$\,GB) & 0.50\,s & from the local server ($4.9$\,GB/s); $\approx20$\,s at $1$\,Gb/s over a network \\
\tabalt Runtime import (ORT-WebGPU + tokenizer) & $\approx$0.3\,s & modules cached by the browser, or vendored \\
Session creation, weights to GPU & $\approx$1.8\,s & includes reading the weights from the local server \\
\tabalt Heads, markers, warm-up & $\approx$0.2\,s & \\
\textbf{Total, repeat load} & \textbf{2.33\,s} & click to ready, phases at $285/2122/2325$\,ms \\
\tabalt Total, first load & 2.5--4.0\,s & fresh browser profile; checks of the published files \\
Question & 110--200\,ms & $152$\,ms mean of $12$ (Table~\ref{tab:battery}) \\
\tabalt \stwo{} generation & 96\,ms/token & $10.5$ tokens/s on the snippets of Table~\ref{tab:dual} \\
\bottomrule
\end{tabularx}
\end{table}

\subsection{Nothing leaves the machine}\label{sec:model-privacy}
The argument for in-browser inference is privacy, so we verified it rather than asserted it. A full session (model load, twelve \sone{} questions, three \stwo{} generations) was recorded with the browser's performance entries. The only hosts contacted are the local server that holds the weights and, in the standalone demo, the public CDN that serves the two runtime modules (ONNX Runtime Web and Transformers.js); the extension of Section~\ref{sec:bis} vendors both, so that even that contact disappears. No state, question, probability or generated text is sent anywhere: inference runs on the user's GPU through WebGPU~\cite{w3c_webgpu}, and the material under test (page source, code, logs) stays in the tab. This is the property that lets a security check read what a hosted service must not be shown.

\section{BIS: the Browser Immune System}\label{sec:bis}
If \aliasweb{} is the brain, BIS is the organ that uses it. BIS is a browser extension (Manifest V3~\cite{chrome_mv3,mdn_webextensions}) that scans every page the user visits with \aliasweb{} running in the browser's own GPU, flags phishing, obfuscated droppers, keyloggers, cryptominers, scareware and fingerprinting while the user navigates, and explains a verdict in plain language when asked.

\subsection{An immune system for the browser}\label{sec:bis-immune}
The name is a design, not a metaphor. A biological immune system has an \emph{innate} response that is fast, inexpensive, always on and relatively unspecific, and an \emph{adaptive} response that is slow, specific and mounted only when the innate layer has raised an alarm~\cite{kephart1994biologically,hofmeyr2000architecture}. The two regimes of \aliasweb{} map onto these tiers exactly. The innate response is a battery of \sone{} questions, each one forward pass with no token decoded, selected by the kind of material observed and run on every page; it costs $0.2$--$0.7$\,s per question on page-sized states and returns a probability per threat class. The adaptive response is \stwo{}, text generated from the same weights, mounted only when the user asks why a page was flagged; it costs seconds and is paid only on positives. Three design rules follow from the measurements of Section~\ref{sec:model}. \emph{The model decides, the browser presents}: the extension contains no detection heuristic of its own, so that every verdict, right or wrong, is attributable to the model or to what it was shown. \emph{Ask per class}: because the readout is conditioned on the question (Figure~\ref{fig:question}), each threat class gets its own directed question. \emph{Spend generation on positives only}: the $6$--$39\times$ cost ratio of Table~\ref{tab:dual} forbids anything else.

\subsection{Architecture}\label{sec:bis-arch}
Figure~\ref{fig:overview} shows the four isolated contexts and what flows between them.

\paragraph{The page: observe without deciding.} A content script injected at \texttt{document\_end} reduces the page to at most $14$ typed \emph{observations}, each a compact text state of at most $1{,}400$ characters (about $350$ tokens): one \texttt{page} observation (domain, title, meta description, headings, visible text), one \texttt{form} observation per form that collects credentials (field types and names, the action URL and whether its host differs from the page's), one \texttt{script-inline} observation per inline script of at least $40$ characters (the code, clipped), one \texttt{script-src} per external script (its URL and the loading page's domain) and one \texttt{iframe} per frame that has a source URL. The script applies no rule beyond this packaging. Two of its choices prove consequential later (\S\ref{sec:bench-errors}, \S\ref{sec:bench-live}): the form observation compares host names rather than origins, and an external script is represented by its URL alone, so that its code is never read.

\paragraph{The service worker: a bus.} MV3's background worker relays each observation batch to the engine and paints the badge. It does not compute. Its one non-obvious design choice is the channel: \texttt{runtime.sendMessage} from the worker did not reliably reach the engine's document in our tests, and the offscreen-document API behaved inconsistently (B15), so requests and verdicts travel through \texttt{chrome.storage} and its \texttt{onChanged} event, which every extension context receives. A request is \texttt{\{url, obs, tabId\}}; a verdict is \texttt{\{level, worst, nQ, ms, findings\}} keyed by tab, plus a \texttt{last} record that wakes the worker to update the badge.

\paragraph{The engine: one model, loaded once.} An engine page (a dedicated tab in the prototype; an offscreen document in a production build) loads \aliasweb{} once and answers every scan. It vendors the two runtime modules and the ONNX Runtime WebGPU WebAssembly binaries ($55$\,MB) inside the extension, because MV3 forbids remote scripts and the runtime bundle does not embed its WebAssembly (B13, B14); the extension's content-security policy adds \texttt{'wasm-unsafe-eval'}. It embeds the tokenizer ($20$\,MB) because Transformers.js hangs, without error, when its local model path is an absolute URL on another origin (B16). Only the $2.44$\,GB graph is fetched from outside the extension, from a local Range-capable server in the prototype. The package is $94$\,MB; the model loads in $2.9$\,s on a repeat load and in about $7$\,s the first time.

\paragraph{The interface.} The badge, the popup and the explanation are described in \S\ref{sec:bis-ux}; they read verdicts from storage and never touch the model.

\subsection{The innate battery}\label{sec:bis-battery}
Table~\ref{tab:policy} is the whole detection logic of BIS: seven directed yes/no questions, one per threat class, and a policy that maps each observation kind to the classes plausible for it. A page observation is asked about phishing and scareware; a credential form about phishing and keyloggers; an inline script about obfuscation, droppers, keyloggers, miners and fingerprinting; an external script about droppers and miners; a third-party iframe about droppers. The policy is what keeps the innate response cheap: the pages of the benchmark, with two to four observations each, cost $4$ to $9$ questions instead of $14$ to $28$. Each question is one \sone{} call of the \texttt{noul} kind on the observation's state; the engine records every answer with $P\geq0.5$ as a finding (\texttt{FLAG}), marks it strong at $P\geq0.8$ (\texttt{STRONG}), and the page verdict is the maximum over findings: green below $0.5$, amber up to $0.8$, red above. The probabilities come from the official head of Eq.~\eqref{eq:head}, not from generated text.

\begin{table}[t]
\centering\bistable\footnotesize
\caption{\textbf{The innate battery}: the policy that selects questions by observation kind, and the seven directed questions. All are yes/no \sone{} questions on the observation's state; a finding is $P\geq0.5$, strong at $P\geq0.8$.}
\label{tab:policy}
\begin{tabularx}{\linewidth}{@{}l Y@{}}
\toprule
\tabhead \textbf{Observation} & \textbf{Classes asked} \\
\midrule
\texttt{page} & phishing, scareware \\
\tabalt \texttt{form} (credentials) & phishing, keylogger \\
\texttt{script-inline} & obfuscation, dropper, keylogger, miner, fingerprint \\
\tabalt \texttt{script-src} & dropper, miner \\
\texttt{iframe} (third party) & dropper \\
\midrule
\tabhead \textbf{Class} & \textbf{Question} \\
\midrule
phishing & Does this page present a login form that sends credentials to a different domain than the one the user believes they are on? \\
\tabalt scareware & Does this page use fake security warnings, invented scan results or false urgency to pressure the user into downloading or paying? \\
keylogger & Does this script capture keyboard or form input from the user and transmit it to a remote endpoint? \\
\tabalt obfuscation & Does this JavaScript contain an obfuscated, encoded or packed payload that gets decoded and executed at runtime? \\
dropper & Does this page load additional executable script code from an external server at runtime? \\
\tabalt miner & Does this script connect to a cryptocurrency mining pool or continuously run hash proof-of-work loops in the browser? \\
fingerprint & Does this script probe many browser, canvas or device properties to build a unique tracking fingerprint of the user? \\
\bottomrule
\end{tabularx}
\end{table}

\subsection{Interface: badge, popup, why}\label{sec:bis-ux}
\begin{figure*}[t]
\centering
\begin{subfigure}[t]{0.235\textwidth}\centering\includegraphics[width=\linewidth]{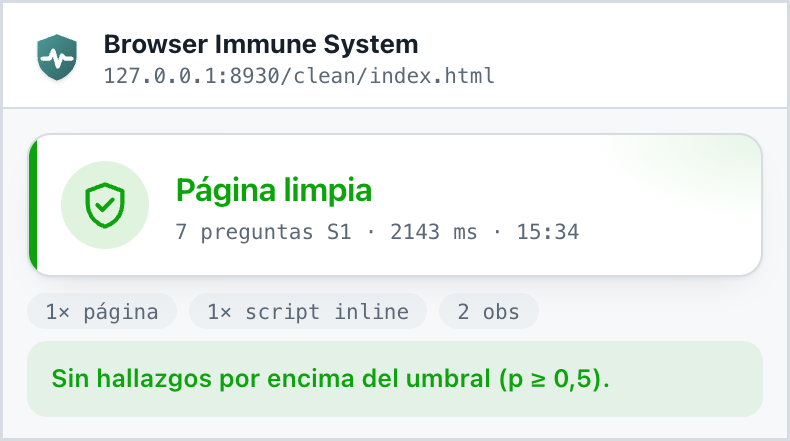}\caption{clean page}\end{subfigure}\hfill
\begin{subfigure}[t]{0.235\textwidth}\centering\includegraphics[width=\linewidth]{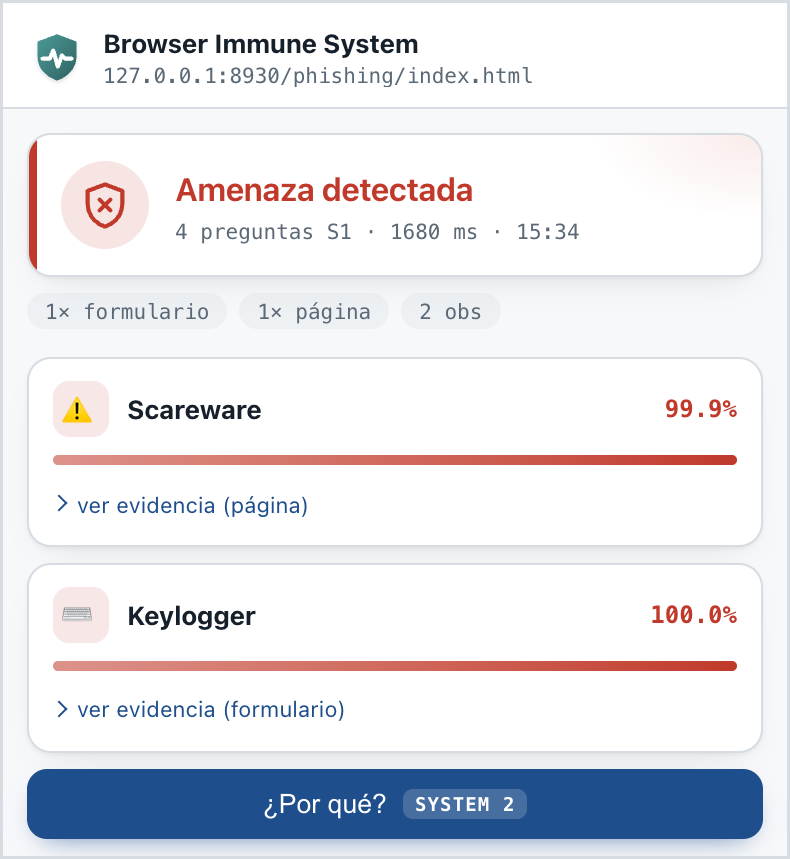}\caption{threat detected}\end{subfigure}\hfill
\begin{subfigure}[t]{0.235\textwidth}\centering\includegraphics[width=\linewidth]{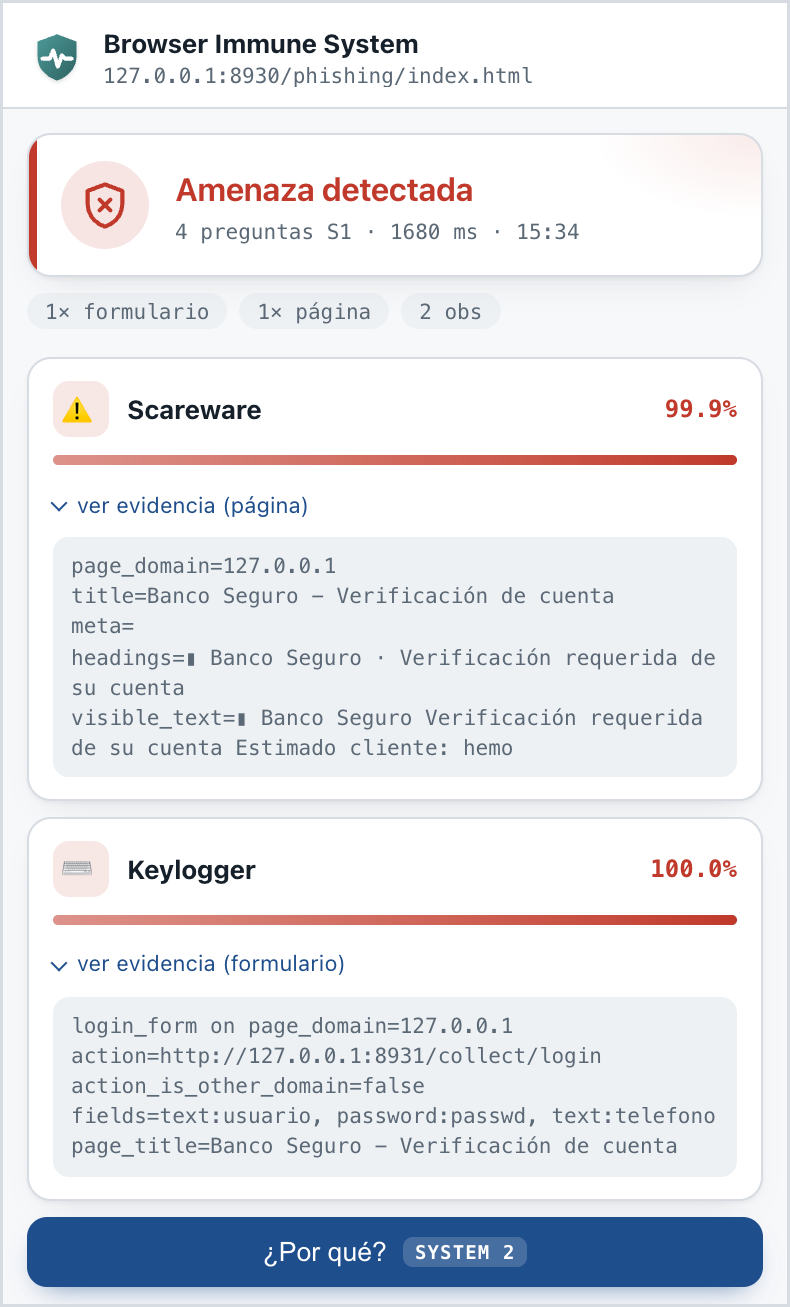}\caption{evidence unfolded}\end{subfigure}\hfill
\begin{subfigure}[t]{0.235\textwidth}\centering\includegraphics[width=\linewidth]{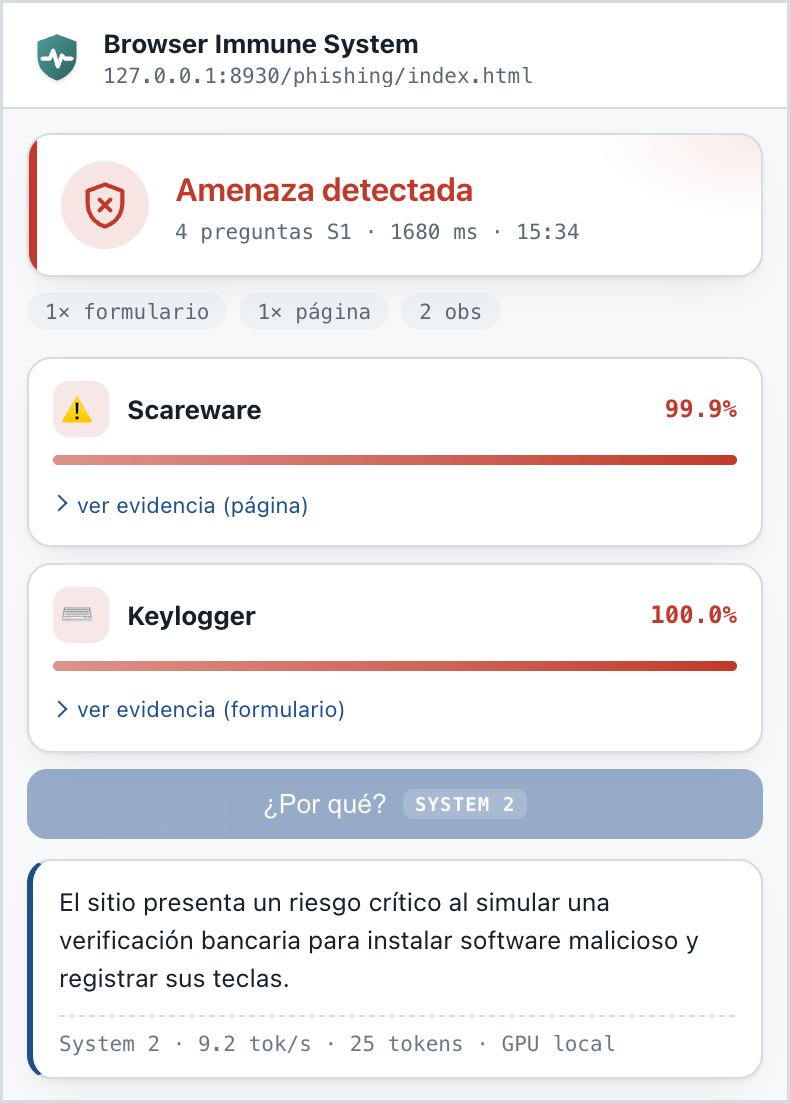}\caption{\stwo{} explanation}\end{subfigure}
\caption{\textbf{The BIS popup on two sites of the benchmark} (Chrome; a later capture than the scored run, so the scan times differ from Table~\ref{tab:results}). (a) A clean page: the verdict, the chips that say what was examined and a note that nothing crossed the threshold. (b) The phishing site: red verdict, one card per finding with the class, the probability bar and the weak/strong mark. (c) The same page with the evidence unfolded: the exact observation that triggered each finding, so that a verdict can be audited. (d) The adaptive response: ``why?'' produces a plain-language explanation from the same weights, with its cost printed ($9$ tokens/s, $25$ tokens, local GPU).}
\label{fig:ux}
\end{figure*}
The interaction is a ramp of three steps with increasing cost to the user (Figure~\ref{fig:ux}). The \emph{badge} costs nothing: the extension icon is painted per tab, a tick in green, the number of findings in amber or red. The \emph{popup} costs one click: a verdict in plain words, chips that list what was examined ($3\times$ inline script, $1\times$ form), and one card per finding with its class, a probability bar, the weak/strong mark and the exact observation that triggered it behind a fold-out, so that the user can audit the evidence rather than trust a box. The \emph{explanation} costs one more click: \stwo{} writes two or three sentences in the user's language from the findings. On the phishing site of Section~\ref{sec:bench} it wrote, in Spanish, ``El sitio presenta un riesgo crítico al simular una verificación bancaria para instalar software malicioso y registrar sus teclas'' (the site poses a critical risk: it imitates a bank verification to install malicious software and record your keystrokes), $25$ tokens at $6$ to $9$ tokens/s across runs; on a green page the button is absent, since there is nothing to explain. The popup refreshes from storage events, so a verdict or an explanation that arrives while it is open appears without a reload. Two generation safeguards were needed in the extension beyond the forced prefix of \S\ref{sec:model-dual}: a directive prompt and a lower temperature ($0.3$), because at $0.7$ the 4-bit model sometimes narrates a pseudo-deliberation as literal text (B18). Finally, the content script exposes each verdict on the page itself, as \texttt{data-bis-verdict} attributes and a \texttt{bis-verdict} event: a testability bridge that lets a harness read verdicts from the page's own context, which is what made the Firefox validation possible (\S\ref{sec:bis-port}) and what an integrator would use in continuous integration.

\subsection{Portability: Chrome and Firefox}\label{sec:bis-port}
All of BIS is standard WebExtension code. The Firefox build differs in one file: Firefox's Manifest V3 implements the background as an event page (\texttt{background.scripts}) and does not accept \texttt{service\_worker}, so a second manifest and a build script produce the Firefox tree and its \texttt{.xpi} ($24$\,MB) with no other change. The hard condition is WebGPU, which the 4-bit graph requires. Table~\ref{tab:webgpu} gives the availability we measured on macOS: the current stable Firefox exposes \texttt{navigator.gpu} but returns no adapter, with or without the fallback adapter and the relevant preferences, and so does Nightly in headless mode; Nightly with a window returns a Metal adapter on the Apple GPU and runs the whole system. Section~\ref{sec:bench-browsers} reports that the verdicts are identical to Chrome's and the compute $5.9\times$ slower. For Chrome, the stable channel ignores the flag that side-loads an unpacked extension, so automated validation uses Chrome for Testing, where the extension also runs headless; a user installs the same directory through the developer mode of the extensions page.

\begin{table}[t]
\centering\bistable\footnotesize
\caption{\textbf{WebGPU availability for the Firefox port}, measured on macOS (Apple M3 Max). Without an adapter the 4-bit graph cannot run, since its embedding gather has no CPU kernel.}
\label{tab:webgpu}
\begin{tabularx}{\linewidth}{@{}l l Y@{}}
\toprule
\tabhead \textbf{Firefox} & \textbf{\texttt{navigator.gpu}} & \textbf{Adapter and outcome} \\
\midrule
159 stable, headless & present & none, also with the fallback adapter and preferences; the engine cannot load the model \\
\tabalt 161 Nightly, headless & present & none \\
161 Nightly, windowed & present & Metal, Apple M3 Max (wgpu): \textbf{runs end to end} \\
\bottomrule
\end{tabularx}
\end{table}

\section{\bisbench{}: Measuring an Immune System}\label{sec:bench}
An immune system is judged by what it catches, what it leaves alone, and how fast. \bisbench{} is the instrument we built to judge BIS. It is small by design and honest about it: eight scenarios, six whose malicious behaviour really executes and two controls built to provoke false alarms, with published labels and metrics that a script recomputes from the raw results.

\subsection{Design: real behaviour, harmless effect}\label{sec:bench-design}
The design principle is \emph{behavioural realism with harmless effect}. Each malicious site performs the conduct of its class (it exfiltrates, connects to a pool, decodes and executes a payload, pressures the user) against local infrastructure that cannot be harmed, and a collector server logs every callback as proof that the signal was present when BIS looked. Table~\ref{tab:scenarios} lists the eight scenarios. Six are threats: a bank notice of unusual access whose login form posts the credentials to the collector; a page whose inline script \texttt{eval}s a hex-escaped second stage that injects a script from a host that does not resolve, and writes a base64-decoded image tag; a webmail login with a keylogger that sends keystrokes to the collector; a miner that opens a WebSocket to a stratum endpoint and runs a capped proof-of-work loop; a scareware page with an invented scan, a countdown and a download button; and a fingerprinting script that reads canvas, WebGL, fonts, user agent and battery and sends the digest. Two are controls: a static cooking blog whose only script writes the year into the footer, and a news page with two advertising iframes and an analytics script, deliberately ``almost suspicious'', because an immune system that panics at ordinary advertising is not usable. The labels (\texttt{expect\_red} per scenario and the classes present per page) live in one file that is the single source of truth for the metrics. Two properties of these fixtures matter for reading the results, and Section~\ref{sec:bench-errors} measures their effect: the eight sites and the collector share one host name and differ only in port, and each malicious script opens with a comment, written for the human who reads the fixture, that names its technique and states that it is harmless.

\begin{table*}[t]
\centering\bistable\footnotesize
\caption{\textbf{\bisbench{} scenarios.} Every threat executes its behaviour for real against a local collector; nothing leaves the machine or damages it. The controls measure specificity.}
\label{tab:scenarios}
\begin{tabularx}{\textwidth}{@{}l l Y l@{}}
\toprule
\tabhead \textbf{Scenario} & \textbf{Label} & \textbf{Behaviour executed} & \textbf{Verified by} \\
\midrule
\texttt{phishing} & threat: phishing & bank notice with a login form that posts to the collector's \texttt{/collect/login} (same host name, another port) & collector \\
\tabalt \texttt{obfuscated} & threat: obfuscation, dropper & \texttt{eval} of a hex-escaped stage that injects a script from a non-resolving host; base64-decoded \texttt{document.write} & execution in page \\
\texttt{keylogger} & threat: keylogger & webmail login; \texttt{keypress} capture sent to \texttt{/collect/keys} & collector \\
\tabalt \texttt{miner} & threat: miner & WebSocket to \texttt{/stratum}; capped FNV proof-of-work loop & collector \\
\texttt{scareware} & threat: scareware & invented ``5 viruses'' scan, countdown, download button & DOM, \texttt{/download} \\
\tabalt \texttt{fingerprint} & threat: fingerprint & canvas, WebGL, fonts, user agent, battery, sent to \texttt{/collect/fp} & collector (full payload) \\
\texttt{clean} & control & static cooking blog; one inline script writes the year & none \\
\tabalt \texttt{adheavy} & control & news page; two advertising iframes and \texttt{analytics.js}, all served by the collector & collector (\texttt{/ad/*} only) \\
\bottomrule
\end{tabularx}
\end{table*}

\subsection{Protocol and metrics}\label{sec:bench-protocol}
A driver navigates Chrome for Testing ($157$, \texttt{headless=new}, Apple GPU) with the extension loaded and the weights served locally, visits the eight sites and harvests each verdict from the extension's own bus, never from the page. Two independent samples are taken per scenario. Because \sone{} is one forward pass with no sampling, the classification is deterministic; the second sample is there to bound the variance of the latency and to verify that the whole pipeline reproduces, not to average accuracy. Four metrics are defined. \textbf{V1, page verdict}: the fraction of scenarios with the expected verdict (red on threats, green on controls); this is the protection metric, what the user sees. \textbf{V2, class, micro}: over (page, class) pairs at threshold $0.5$, a true positive is a labelled class detected, a false negative a labelled class missed, and a false positive a class detected on a page not labelled with it, \emph{including cross-attributions} (a keylogger finding on the phishing page counts as a false positive). \textbf{V3, class, macro}: the unweighted mean of per-class precision and recall. \textbf{V4, latency}: \emph{scan}, the milliseconds spent in the \sone{} battery, and \emph{wall}, the seconds from navigation to visible verdict. All four are recomputed by one script from the result and label files; nothing is typed by hand. The same harness runs in Firefox (\S\ref{sec:bench-browsers}).

\subsection{Results}\label{sec:bench-results}
\begin{figure*}[t]
\centering
\includegraphics[width=\textwidth]{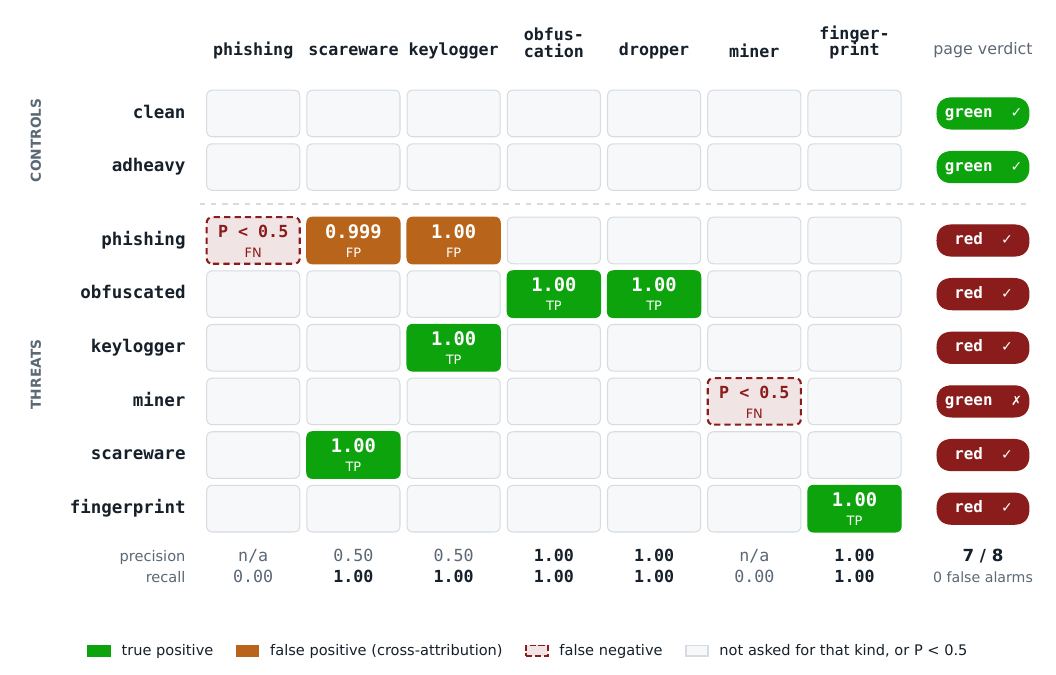}
\caption{\textbf{\bisbench{} detection matrix} ($n=2$, identical in both samples). Rows are scenarios, columns the classes of the battery, cells the maximum $P(\text{yes})$ of the class on the page. Green: detected and labelled (TP). Ochre: detected without label (FP, the two cross-attributions on the phishing page, which still ends red). Dashed red: labelled and not detected (FN, the class phishing and the miner). Pale cells: classes not asked for that observation kind, or below $0.5$. The bottom rows give precision and recall per class; the right column the page verdict.}
\label{fig:matrix}
\end{figure*}

Table~\ref{tab:results} gives the per-scenario outcome and Figure~\ref{fig:matrix} the detection matrix; Table~\ref{tab:metrics} the metrics. Five of the six threats are red with the worst finding at $P\geq0.999$, both controls are green with no finding at all, and the miner is the one escape. At class level there are five true positives, two false negatives (the class \emph{phishing} and the miner) and two false positives, both cross-attributions on the phishing page, which the model flags as scareware ($0.999$) and keylogger ($1.000$) rather than as phishing. Precision is $1.00$ for obfuscation, dropper and fingerprint on these pages, and for every finding the popup shows the observation that triggered it. The scan costs $4$ to $9$ questions and $0.9$ to $6.4$\,s per page, $2.62$\,s on average; the longest case is the obfuscated page, whose nine questions on long script states take $6.4$\,s.

\begin{table*}[t]
\centering\bistable\footnotesize\setlength{\tabcolsep}{4pt}
\caption{\textbf{\bisbench{} results}, Chrome for Testing, $n=2$ independent samples per scenario. The verdict and every probability were identical in both samples; scan is the \sone{} battery time (range over the two samples), wall the navigation-to-verdict time of each sample.}
\label{tab:results}
\begin{tabular}{@{}llllrlrl@{}}
\toprule
\tabhead \textbf{Scenario} & \textbf{Expected} & \textbf{Verdict (2/2)} & \textbf{Worst $P$} & \textbf{Findings ($P$)} & \textbf{$n_Q$} & \textbf{scan ms (range)} & \textbf{wall s} \\
\midrule
\texttt{clean} & green & \cellcolor{fig_s1!12}green \ding{51} & 0.00 & none & 7 & 1\,221 (1\,208--1\,234) & 4.1 / 2.0 \\
\tabalt \texttt{adheavy} & green & \cellcolor{fig_s1!12}green \ding{51} & 0.00 & none & 6 & 1\,110 (1\,096--1\,123) & 2.0 / 2.0 \\
\texttt{phishing} & red & \cellcolor{fig_s1!12}red \ding{51} & 1.00 & scareware (0.999), keylogger (1.00) & 4 & 927 (926--928) & 2.0 / 2.0 \\
\tabalt \texttt{obfuscated} & red & \cellcolor{fig_s1!12}red \ding{51} & 1.00 & obfuscation (1.00), dropper (1.00) & 9 & 6\,428 (6\,408--6\,447) & 8.0 / 8.0 \\
\texttt{keylogger} & red & \cellcolor{fig_s1!12}red \ding{51} & 1.00 & keylogger (1.00) & 9 & 2\,164 (2\,163--2\,165) & 4.0 / 4.0 \\
\tabalt \texttt{miner} & red & \cellcolor{fig_warning!12}green \ding{55} & 0.00 & none & 7 & 3\,160 (3\,139--3\,181) & 4.0 / 4.0 \\
\texttt{scareware} & red & \cellcolor{fig_s1!12}red \ding{51} & 1.00 & scareware (1.00) & 7 & 1\,784 (1\,764--1\,803) & 4.0 / 4.0 \\
\tabalt \texttt{fingerprint} & red & \cellcolor{fig_s1!12}red \ding{51} & 1.00 & fingerprint (1.00) & 7 & 4\,152 (4\,130--4\,173) & 6.0 / 6.0 \\
\bottomrule
\end{tabular}
\end{table*}

\begin{table}[t]
\centering\bistable\footnotesize
\caption{\textbf{\bisbench{} metrics}, recomputed from the results and labels.}
\label{tab:metrics}
\begin{tabularx}{\linewidth}{@{}l l Y@{}}
\toprule
\tabhead \textbf{Metric} & \textbf{Value} & \textbf{Reading} \\
\midrule
V1 threats in red & $5/6=0.833$ & the escape is the miner \\
\tabalt V1 controls in green & $2/2=1.000$ & no false positive, including the ``almost suspicious'' control \\
V1 accuracy & $7/8=0.875$ & the verdict the user sees \\
\tabalt V2 micro P / R / F1 & $0.714$ / $0.714$ / $0.714$ & TP $5$, FP $2$, FN $2$; cross-attributions count as FP \\
V3 macro P / R & $0.800$ / $0.714$ & when it accuses, the behaviour is almost always there \\
\tabalt V4 scan, mean & $2.62$\,s per page & $4$--$9$ questions; wall $2$--$8$\,s \\
\bottomrule
\end{tabularx}
\end{table}

\paragraph{Reproducibility.} All $16$ verdicts, and every probability behind them, are identical between the two samples; scan times differ by at most $2.5\%$ between samples (worst case the ad-heavy control, $1{,}096$ against $1{,}123$\,ms). The model loaded in $2.9$\,s in this run (about $7$\,s on a first load). The full run of $2\times8$ navigations took $864$\,s, dominated by the driver's two-second polling, not by BIS. These properties are what make \bisbench{} a regression harness rather than a one-off measurement: a change to the policy, a question or the model can be re-measured in minutes against comparable numbers (Appendix~\ref{app:repro}). The diagnostic run of Section~\ref{sec:bench-errors}, days later, visited the eight sites once more and returned the same eight verdicts.

\subsection{Cross-browser: identical verdicts, slower compute}\label{sec:bench-browsers}
\begin{figure}[t]
\centering
\includegraphics[width=\linewidth]{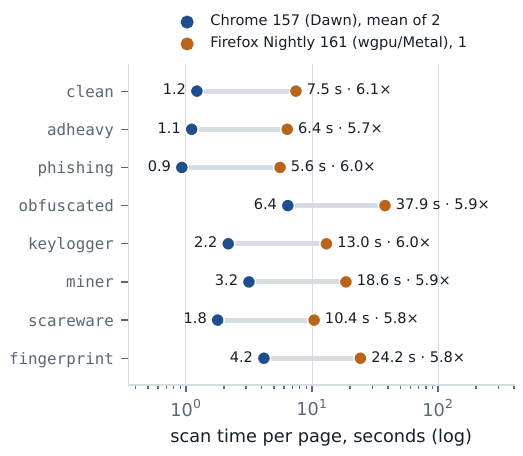}
\caption{\textbf{Scan time per scenario, Chrome against Firefox}, same GPU and weights. Chrome for Testing $157$ (Dawn, mean of two samples) and Firefox Nightly $161$ with a window (wgpu on Metal, one sample); the ratio is printed beside each pair. Verdicts and probabilities are identical in both browsers.}
\label{fig:browsers}
\end{figure}

The Firefox port (\S\ref{sec:bis-port}) was validated with the same eight scenarios, one sample, in Firefox Nightly $161$ with a window, driven through geckodriver with the extension installed as a temporary add-on and verdicts read from the testability bridge, since the driver refuses to inspect privileged extension pages. All eight verdicts, and all probabilities, are identical to Chrome's (Figure~\ref{fig:browsers}). The compute is not: WebGPU reaches the GPU through \texttt{wgpu} and Metal rather than Chrome's Dawn, and every scenario takes $5.7$--$6.1\times$ longer ($2.20$\,s per question against $0.37$; the obfuscated page $37.9$ against $6.4$\,s). The uniformity of the ratio says that the extra cost is the compute pipeline, not pathological cases, and the identical verdicts say that \sone{} is a property of the weights and the states, not of the runtime. The model loaded in $4.6$\,s; the whole sample took $139$\,s.

\subsection{Error analysis: one variable at a time}\label{sec:bench-errors}
Because the extension holds no rule of its own and \sone{} does not sample, an error can be localised by experiment: take an observation exactly as the engine received it, change one thing, and ask again. Table~\ref{tab:diag} lists the diagnostics run this way after the scored campaign, without touching a fixture or the extension.

\begin{table*}[t]
\centering\bistable\footnotesize
\caption{\textbf{Single-variable diagnostics on \bisbench{}}, outside the scored run. Each row takes an observation exactly as the engine received it, changes one thing and asks the same question again; D5 and D6 load a page through the real pipeline. \sone{} does not sample, so one measurement per state is exact.}
\label{tab:diag}
\begin{tabularx}{\textwidth}{@{}l Y l l l@{}}
\toprule
\tabhead & \textbf{What is changed} & \textbf{Question} & \textbf{As scored} & \textbf{Changed} \\
\midrule
D1 & Miner script: its two comments removed & miner & $0.0001$ & $0.9999$ \\
\tabalt D2 & Miner script without comments: the line \texttt{// harmless test code} prepended & miner & $0.9999$ & $0.0070$ \\
D3 & Keylogger, obfuscated and fingerprinting scripts: the same line prepended & labelled class & $\geq0.9999$ & $\geq0.9999$ \\
\tabalt D4 & Phishing form: \texttt{action\_is\_other\_domain} set from \texttt{false} to \texttt{true} & phishing & $0.0000$ & $1.0000$ \\
D5 & Phishing page served under another host name, end to end & page verdict & red, phishing not found & red, phishing $1.00$ \\
\tabalt D6 & A legitimate login page whose form posts to its own origin, end to end & keylogger, on the form & no such control & $0.9999$, page red \\
D7 & Ad-heavy control: the advertising host given another name & dropper, on the script & $0.437$ & $0.934$ \\
\tabalt D8 & Keylogger, obfuscated and fingerprinting scripts: self-description removed & labelled class & $\geq0.9999$ & $\geq0.9999$ \\
\tabalt & & other classes & $<0.06$ & miner $0.992$; dropper $0.989$, $0.785$ \\
\bottomrule
\end{tabularx}
\end{table*}

\emph{The miner} is not missed for lack of evidence. Its script is $699$ characters long and reached the model whole, stratum socket and proof-of-work loop included. It is missed because the script opens with the fixture's own comment, which says in plain words that this is a harmless miner pattern with a nonexistent pool and a bounded loop, and the model takes the page at its word: the miner question returns $0.0001$ with the comments and $0.9999$ without them (D1). A neutral comment in their place changes nothing ($0.9998$); a generic line that claims harmlessness brings the answer back down to $0.007$ (D2), although the same line leaves the three other script detections where they were (D3) and loses most of its effect when written as a block comment ($0.98$). \emph{The class phishing} is missed because of one field of the observation. The content script compares host names, and in \bisbench{} the collector listens on the site's own host name, on another port, so the form state read \texttt{action\_is\_other\_domain=false} and the model answered the question it was shown. With the flag set to \texttt{true} and nothing else changed the answer goes from $0.0000$ to $1.0000$ (D4), and the same page served under another host name ends red with phishing at $1.00$ through the real pipeline (D5). Neither error lies in the head and neither is a quantisation artefact: the two samples and the two browsers return the same probabilities.

The same instrument shows what the scored run could not. \emph{Two rules of the policy are not specific.} The keylogger question, asked of a credential form, answers yes to any form that collects a password: a legitimate login page whose form posts to its own origin ends red at $0.9999$ (D6). That is where the keylogger finding on the phishing page comes from, and no control of \bisbench{} contains a credential form that would have exposed it. The dropper question, asked of an external script, turns on the name of the host: the analytics script of the ad-heavy control scores $0.437$, just under the threshold, when it comes from the page's own host name as in the benchmark, and $0.93$ when the advertising host is given another name (D7); a same-origin script on a named site scores $0.9999$ and a common library on a public content network $0.69$. \emph{The fixtures' comments move answers in both directions.} With the self-descriptions removed from the three other script fixtures every labelled class stays above $0.9999$, but three findings appear that the comments had suppressed: miner on the obfuscated script, and dropper on the keylogger and on the fingerprinting script (D8).

Read together, the diagnostics bound what Table~\ref{tab:results} supports. The page verdicts do not depend on the comments, except for the miner, where the comment is the cause of the miss, and the detections are robust to a claim of harmlessness in three cases out of four. The class-level precision of Table~\ref{tab:metrics} and the two green controls, by contrast, are properties of version $1.0$ of the fixtures: they owe something to the shared host name, to the absence of a login form among the controls and to the comments. The remedies sit in the benchmark, the observation and the policy, not in the model: fixtures that do not describe themselves and that use distinct host names, a control with a legitimate login, a comparison of origins in the form observation, the keylogger question asked of scripts only, and a dropper question conditioned on more than the presence of an external script.

\subsection{Beyond the benchmark: eleven real public pages}\label{sec:bench-live}
\bisbench{} is synthetic by design. To see where it stops we pointed the same extension, unchanged, at eleven public pages that exist to exercise browser protections or that are plainly benign (Table~\ref{tab:live}): the three test pages of a blocklist service, a public phishing trap, three browser-fingerprinting demonstrations, two deliberately vulnerable teaching applications, and two clean controls, an encyclopedia article and a one-paragraph placeholder page. It is a probe, not a benchmark: one sample, no labels beyond the nature of each page, no claim of coverage.

\begin{table*}[t]
\centering\bistable\footnotesize
\caption{\textbf{A probe on eleven real public pages}, one sample, Chrome for Testing, same extension and weights as \bisbench{}; verdicts read through the testability bridge. A page's verdict here says how the current battery behaves on it, not what the page is: every red in this table comes from the dropper question on external scripts or the keylogger question on a form.}
\label{tab:live}
\begin{tabularx}{\textwidth}{@{}l l l Y r r@{}}
\toprule
\tabhead \textbf{Page} & \textbf{Nature} & \textbf{Verdict} & \textbf{Findings ($P$)} & \textbf{$n_Q$} & \textbf{scan s} \\
\midrule
\texttt{testsafebrowsing.appspot.com} (3 pages) & blocklist test pages & green & none: the content is innocuous, the signal is the URL & 2 & 0.3--1.2 \\
\tabalt \texttt{itisatrap.org} & phishing trap & none & the page did not finish loading in headless mode & -- & -- \\
\texttt{browserleaks.com/javascript} & fingerprinting demonstration & red, other class & keylogger on a form ($1.00$), dropper ($0.56$) & 25 & 5.6 \\
\tabalt \texttt{amiunique.org} & fingerprinting demonstration & red, other class & dropper on $13$ external scripts ($0.83$--$1.00$) & 28 & 5.0 \\
\texttt{coveryourtracks.eff.org} & fingerprinting demonstration & red, other class & dropper on $3$ external scripts ($0.53$--$0.93$) & 16 & 3.3 \\
\tabalt \texttt{google-gruyere.appspot.com} & vulnerable teaching application & green & none & 7 & 3.1 \\
\texttt{xss-game.appspot.com} & vulnerable teaching application & red, false positive & dropper ($1.00$) & 4 & 1.2 \\
\tabalt \texttt{en.wikipedia.org/wiki/Security} & clean control & red, false positive & dropper ($1.00$) & 24 & 13.8 \\
\texttt{example.com} & clean control & red, false positive & dropper ($1.00$) & 4 & 1.2 \\
\bottomrule
\end{tabularx}
\end{table*}

Three things follow. \emph{The dropper question is not calibrated for the real web.} ``Loads executable code from an external server'' describes almost any page: both clean controls end red on it, each for a single external script at $P=1.00$, and so does one of the teaching applications, whose front page does nothing malicious. \emph{What BIS reads decides what it can find.} The three fingerprinting pages are red, but through the dropper question and, on one of them, the keylogger question on a form; the fingerprint question, which is asked of inline code only, never fires, and code that a page loads from external files is represented by its URL alone and never read. The test pages of the blocklist service are innocuous in content, their signal lives in the reputation of the URL, and BIS, which reads content, finds nothing: they contrast the two approaches rather than score this one. \emph{Cost scales with the page.} The scan ranges from $2$ questions and $0.3$\,s to $28$ questions, and reaches $13.8$\,s on the encyclopedia article ($24$ questions on six observations), against $0.9$--$6.4$\,s on the synthetic sites.

The probe and the diagnostics of Section~\ref{sec:bench-errors} agree on where the open problem is. The innate battery detects the behaviours it was built for; two of its seven questions, dropper on external scripts and keylogger on forms, answer yes on ordinary pages. Specificity on the real web, measured on a ranked list of popular sites, is the calibration this system still owes.

\subsection{What it costs to run}\label{sec:bench-cost}
Table~\ref{tab:cost} gives the measured cost of a campaign. The one-time setup (Chrome for Testing, the WebAssembly binaries and the tokenizer) takes $10$--$12$ minutes; a run starts in about $20$\,s; a sample of eight sites takes $7.2$ minutes, of which BIS itself accounts for $21$\,s of scanning; the two-sample campaign took $15$ minutes. The runbook is in Appendix~\ref{app:repro}.

\begin{table}[t]
\centering\bistable\footnotesize
\caption{\textbf{Cost of a \bisbench{} campaign}, measured.}
\label{tab:cost}
\begin{tabularx}{\linewidth}{@{}l l Y@{}}
\toprule
\tabhead \textbf{Step} & \textbf{Time} & \textbf{Note} \\
\midrule
one-time setup & 10--12 min & Chrome for Testing, WebAssembly binaries, tokenizer \\
\tabalt start of a run & $\approx$20 s & three local servers, Chrome, extension registration \\
first navigation & +3--7 s & model load, repeat or first \\
\tabalt one sample (8 sites) & $\approx$7.2 min & driver polling dominates; scanning is $2.62$\,s per site \\
two samples & $\approx$15 min ($864$\,s) & verdicts deterministic: two samples bound the latency \\
\bottomrule
\end{tabularx}
\end{table}

\section{Discussion}\label{sec:discussion}
\paragraph{What BIS demonstrates.} Four results survive the controls. First, \sone{} runs in the browser \emph{as designed}: with the official marker ids, the complete head and the final normalisation, the 4-bit graph answers the validation battery as expected in about $150$\,ms per question, in the bf16 model's decision direction on the prompt we compared, and the same loaded weights generate correct diagnoses at $10.5$ tokens/s (Section~\ref{sec:model}). This closes the gap the prior study left open, and it does so on a derivative with a context-category sensor rather than on the public release. Second, the innate/adaptive division holds up as an architecture: a battery of directed \sone{} questions screens a page of the benchmark in $0.9$--$6.4$\,s, generation is paid only when the user asks why, and the extension carries no detection heuristic of its own, so that each of its errors is attributable (Section~\ref{sec:bis}). Third, on \bisbench{} the system returns the expected page verdict on $7/8$ sites, and both of its errors were localised by experiment to one line of what the model was shown, a comment in a fixture and a field of an observation; with either corrected, the missing answer is $1.00$ (Section~\ref{sec:bench-errors}). That an error can be traced this way is the practical content of the rule that the model decides and the browser presents. Fourth, the result is deterministic and portable: $16/16$ verdicts identical across samples, $8/8$ identical across Chrome and Firefox with the same probabilities, which turns the benchmark into a regression harness and the port into a one-file change.

\paragraph{What the evidence does not support.}\label{sec:limits} We gather every limit in one place. \emph{Specificity.} This is the main one. The absence of false positives on \bisbench{} holds for its two controls as built: the ad-heavy control passes at $0.437$ because its advertising host shares the page's host name, and no control contains a credential form. The diagnostics and the probe on real pages agree that two rules of the policy, dropper on external scripts and keylogger on forms, answer yes on ordinary pages: both clean pages of the probe and a plain login page ended red. Until those rules are recalibrated and false positives are measured on popular benign sites, BIS is a detector of behaviours with a known over-sensitivity, not a filter ready to deploy. \emph{Coverage.} \bisbench{} has eight single-page synthetic scenarios, no active evasion, no mixing of signals and no single-page applications. Its fixtures describe themselves in comments, which moves answers in both directions, and its class-level figures are those of version $1.0$ of the fixtures (Section~\ref{sec:bench-errors}). It is not comparable with the corpora of the classic detectors of Section~\ref{sec:related}, whose authors measured on thousands of real pages. The prior study's $172$-question set shows what sample sizes can and cannot detect; ours gives no statistical power on accuracy, only on reproducibility. \emph{Parity.} The agreement of the 4-bit readout with the bf16 reference was checked on one prompt; how often the official head's answers change under quantisation is not measured. \emph{What is observed.} An external script is represented by its URL and its code is never read, so a page whose logic lives in external files, as on most of the real web, is judged on where its code comes from, not on what it does. Inline scripts are read up to $1{,}400$ characters, a bound that no fixture reaches and whose effect is therefore unmeasured. \emph{Semantics of the sensor.} The \texttt{choice} and \texttt{score} kinds speak the language of the checkpoint's training (campaign context categories), not a weakness taxonomy or a calibrated severity; BIS uses only yes/no questions for that reason. \emph{WebGPU dependency.} The 4-bit graph has no CPU path; on macOS only Firefox Nightly with a window runs it, $5.9\times$ slower than Chrome; Safari and Windows are unmeasured. \emph{Distribution.} The $2.44$\,GB graph is served from a local server; packaging for a store (a stable offscreen engine, weights by range requests from a content network, a signed Firefox package) is pending. \emph{Hardware.} Every number comes from one Apple M3 Max with $128$\,GB; there is no measurement on a modest integrated GPU, where the $2.44$\,GB residency and the dispatch-bound generation will bite first. \emph{Threat model.} BIS judges what the content script shows it. An adversary who keeps the malicious logic outside the observed material, past the clip, in an external file, in a deferred load, in a frame the policy does not ask about, evades by construction. Text inside the observed material is a second channel: one comment that claims harmlessness took the miner answer from $1.00$ to $0.00$, by accident in a fixture and again on purpose with a generic line, while the same line left three other detections unmoved. We have not tested this systematically, and the sanitiser only prevents forging the option markers. The \stwo{} explanation reads page-derived text and is exposed to prompt injection as any generator is. These are open problems for an immune system whose antigens are written by the attacker.

\paragraph{Lessons that generalise.} Three obstacles of the bring-up carry beyond this model. \emph{The trained form must be honoured exactly.} The three costliest bugs (B1, B3, B9 in Appendix~\ref{app:engineering}) were departures from it: a misnamed dimension field that silenced the projections, a pre-norm state fed to a post-norm head, a special token passed through a sanitiser; once the form was respected (ids, explicit norm, softmax over options) the 4-bit graph reproduced the bf16 decision. \emph{Exported hidden states must be checked for normalisation.} Two builds of the same model from the same tool exposed different nodes; a $10$\,KB file resolved a failure that looked like quantisation damage and was not. \emph{Determinism is testability.} A readout without sampling gives identical verdicts across samples and browsers, so two samples suffice for latency and any difference between runs is a real change, which is what a regression harness needs. A fourth lesson is organisational: keeping the browser side free of heuristics made every error localisable in minutes. A fifth concerns benchmarks: comments written for the human reader of a fixture are read by the model too; ours named each technique and declared it harmless, and moved answers in both directions.

\paragraph{Head against logits.} The prior study compared the recovered head with the next-token logits and found a tie on yes/no and an advantage for the head on choice and severity. We did not repeat that comparison on \bisbench{}; our contribution is to run the official head on the official interface, and the controlled comparison of the two readouts on the same pages is the natural next experiment.

\paragraph{Extensions.} In order of expected return: recalibrate the two rules that are not specific (ask the keylogger question of scripts only; condition the dropper question on more than the presence of an external script, or replace it) and measure false positives on a ranked list of popular sites; release version $1.1$ of the benchmark, with fixtures that do not describe themselves, distinct host names and a control with a legitimate login, and compare origins in the form observation, after which the diagnostics predict eight expected page verdicts on the current scenarios; read the code of external scripts, and long scripts block by block; test text written to move the readout as an attack in its own right; calibrate severity on this graph with the prior work's \texttt{head\_score} recipe, which gives meaning to the \texttt{score} kind; extend the benchmark to sixteen scenarios with staged obfuscation, deferred loading and single-page applications; compare BIS with a blocklist service and with a classic in-browser classifier on one shared corpus; package for distribution; and measure the floor of viability on an 8\,GB laptop and on an x86 integrated GPU through Dawn on Direct3D~12.

\section{Conclusion}\label{sec:conclusion}
We presented \aliasweb{}, a cybersecurity transformer compiled for the browser, and the Browser Immune System built on it. \aliasweb{} is the first deployment of the typed decision mode of~\cite{mayoralvilches2026transformer} outside a server that uses the model's official pieces, the marker tokens, the complete pointer head and the final normalisation, and it shows that a cybersecurity transformer quantised to 4 bits can answer a typed security question through its trained readout in $152$\,ms on a laptop GPU, in the direction of its bf16 original where the two were compared, and write explanations from the same weights at $10.5$ tokens/s. BIS turns those two regimes into an innate response that screens every page with one directed question per threat class and an adaptive response that explains on request; on \bisbench{}, a benchmark of behaviourally real but harmless malicious sites, it returns the expected verdict on $7/8$ pages, with identical verdicts across samples and across two browsers, and single-variable diagnostics trace both of its errors to what the model was shown. The prior edge study established that \sone{} answers survive compression; this paper runs the readout \emph{as trained} on the compressed graph and shows that a browser can screen navigation with it on a machine that shares nothing with anyone. It also establishes where the work now lies: the same diagnostics and a first pass over real pages show that two of the seven questions answer yes on ordinary scripts and forms, and that a line of text inside a page can talk the readout out of a detection. Like the model it starts from, BIS is a starting point: each error we found was localised to a question or to a line of an observation, by a procedure that reruns in about a minute.

\section*{Acknowledgements}
This work has been funded by the European Innovation Council (GA 101161136).

\bibliography{bis-bibliography}

\onecolumn
\appendix

\section{Reproduction}\label{app:repro}
\paragraph{Building \aliasweb{}.} The graph is built on Linux x86\_64 from the fused checkpoint with the recipe of the prior brief~\cite{worrel2026bestbuild}, adding the two readout files:
\begin{lstlisting}
pip install onnxruntime-genai==0.18.0 transformers numpy onnx onnx-ir safetensors
# build.py: the brief's 20-line wrapper around onnxruntime_genai.models.builder; it relabels the
# text-only Qwen3_5ForCausalLM checkpoint so that the 0.18 builder dispatches on it
python build.py -i ./alias-web-fused -o ./model -p int4 -e webgpu \
  --extra_options include_hidden_states=true exclude_mtp=true      # 16 s, 2.44 GB
# heads/head.bin : the complete pointer head, fp16, runner header (2 622 545 B)
# heads/norm.bin : model.norm.weight, fp32 (10 289 B)
\end{lstlisting}
The bundle (\texttt{model/}, \texttt{heads/}, the vendored runner, \texttt{aliasweb.mjs}, a demo page and a Range-capable server) is published as \texttt{alias3-web} at \href{\aliasweburl}{\texttt{huggingface.co/\allowbreak aliasrobotics/\allowbreak alias3-web}} and is served locally; a page loads it with \texttt{AliasWeb.load(\{base:"./"\})} and asks \texttt{ask(\{kind:"noul", state, q\})}.

\paragraph{Installing BIS.} Fetch the vendored binaries once (\texttt{fetch-vendor.sh}: the ONNX Runtime WebAssembly files and the tokenizer, $75$\,MB), start the weight server (\texttt{serve.py 8924}), and in Chrome open \texttt{chrome://extensions}, enable developer mode and load the \texttt{bis-extension/} directory unpacked. For Firefox, \texttt{build-firefox.sh} writes \texttt{dist/bis-firefox.xpi}, loaded as a temporary add-on from \texttt{about:debugging} in Nightly. The first page visited pays the model load ($2.9$--$7$\,s); each page of the benchmark afterwards costs $4$--$9$ questions, and real pages up to $28$ (Section~\ref{sec:bench-live}).

\paragraph{Running \bisbench{}.}
\begin{lstlisting}
# one-time (10-12 min): Chrome for Testing + vendored binaries
npx -y @puppeteer/browsers install chrome@stable --path ~/.cache/puppeteer
cd bis-extension && ./fetch-vendor.sh
# each run (~20 s): weights, the eight sites, the harmless collector, Chrome
python3 serve.py 8924 &  python3 bench_server.py 8930 &  python3 collector_server.py 8931 &
"$CFT" --headless=new --remote-debugging-port=9222 --user-data-dir=/tmp/bis-profile \
  --load-extension=bis-extension --disable-extensions-except=bis-extension &
BIS_SAMPLES=2 node bench.mjs        # verdicts from the extension's bus -> results2.json
python3 metrics.py results2.json    # V1-V4 recomputed from results + labels.json
node diagnostics.mjs                # Section 6.5: one variable at a time -> results-diagnostics.json
node live.mjs                       # Section 6.6: the probe on real public pages -> results-live.json
\end{lstlisting}
The Firefox validation installs the \texttt{.xpi} through geckodriver and Selenium in Nightly with a window (the only configuration with a WebGPU adapter on macOS), reads verdicts from the page's \texttt{data-bis-verdict-json} attribute, and writes \texttt{results-firefox.json}; it takes about $4$ minutes of setup and $2.5$ minutes to run. The diagnostics script reads the seven questions from the extension's source, sends the eight sites through the real pipeline, and asks the engine again about the exact observation states with one change each; it serves its two auxiliary pages itself and runs in about a minute.

\paragraph{Provenance of the prior-work numbers.} The prior study's research archive contains the $172$-question decision test set, the per-question results of every build, the benchmark file from which its figures and tables are generated, the pre-registration and final tables of the per-class-question study ($161$ real items, $117$ positive), and a re-run of its recipe by an independent agent: the rebuilt graph has the same SHA-256 as the published one, $187$ of $621$ initialisers differ on Apple Silicon (about $0.02\%$ of the int4 bytes and $5$--$8\%$ of the rotary-table bytes), and the sampled tensors are identical on Linux x86. Table~\ref{tab:prior} and Figure~\ref{fig:latency}a take their timing values from that archive's measurement files.

\section{Engineering record}\label{app:engineering}
Tables~\ref{tab:mods}, \ref{tab:bugs} and~\ref{tab:ffbugs} record what had to be changed and what broke between the research checkpoint and the working extension. They are kept because several entries cost days, two generalise (Section~\ref{sec:discussion}), and a reader reproducing the work will meet most of them.

\begin{table}[ht]
\centering\bistable\footnotesize
\caption{\textbf{Modifications} with respect to the prior recipe and the default behaviour of the prior runner.}
\label{tab:mods}
\begin{tabularx}{\textwidth}{@{}l Y Y@{}}
\toprule
\tabhead \textbf{\#} & \textbf{Modification} & \textbf{Reason} \\
\midrule
M1 & \texttt{norm.bin} (new, $10$\,KB): the final RMSNorm weight, applied in JavaScript before the head & the graph exports pre-norm hidden states; the head was trained on post-norm states; without it the scores saturate at $\mathrm{e}^{-52}$ \\
\tabalt M2 & marker ids read from the tokenizer's \texttt{added\_tokens} & Transformers.js exposes no token-to-id conversion \\
M3 & own encoder: one causal row with the official ids and the \texttt{<|x|>} sanitiser & the runner's template writes markers as plain text; the official interface uses ids that the sanitiser must protect \\
\tabalt M4 & tokenizer duplicated under \texttt{model/tok/}, heads beside the graph & path conventions of the runner in local mode \\
M5 & weight server with HTTP Range support & the $2.44$\,GB file is fetched by ranges; a plain static server cannot serve it \\
\tabalt M6 & import map and vendored runner instead of a package build & self-contained bundle; the runner's bare imports resolved to pinned modules \\
M7 & WebGPU-only session, no CPU fallback, placement checks off & the quantised gather and the DeltaNet kernels exist only on the WebGPU provider; placement was verified once against this graph \\
\tabalt M8 & own \texttt{generate()} with the forced \texttt{</think>} prefix & the 4-bit checkpoint deliberates without closing (B8) \\
M9 & official scale $1/\sqrt{256}$ and softmax over options for every kind, yes/no included & the runner applies a sigmoid to a single recovered-head score; the trained form is a softmax over options \\
\tabalt M10 & \texttt{head.bin} regenerated from the complete head, temperature $1.0$ in the header & the public file is truncated; the checkpoint's temperature is stored with it \\
\bottomrule
\end{tabularx}
\end{table}

\begin{table}[ht]
\centering\bistable\footnotesize
\caption{\textbf{Bring-up bugs}, from the model in a page (B1--B11) to the extension in Chrome (B12--B19).}
\label{tab:bugs}
\begin{tabularx}{\textwidth}{@{}l Y Y Y@{}}
\toprule
\tabhead \textbf{\#} & \textbf{Symptom} & \textbf{Root cause} & \textbf{Fix} \\
\midrule
B1 & identical raw scores ($-3\cdot10^{-4}$) for every option & the head's dimension field is \texttt{hiddenSize}, not \texttt{hidden}: the projections received an undefined size and returned only their biases & one line; scores went from constant to $\pm2$--$4$ \\
\tabalt B2 & own port of the runner: NaNs in $398/2560$ hidden dimensions & dynamic DeltaNet cache dimensions ($[1,8192,3]$) mapped to the attention head dimension & keep the vendored runner \\
B3 & official head saturates ($\mathrm{e}^{-52}$), meaningless winner & pre-norm state read by a post-norm head; magnitudes $5\times$ smaller than in training & M1; verified against bf16 PyTorch \\
\tabalt B4 & no token-to-id conversion & API absent in Transformers.js v3 & M2 \\
B5 & cannot debug on CPU & the quantised embedding gather has no CPU kernel & pairwise isolation: same graph, vendored runner (works) against own runner (fails) \\
\tabalt B6 & public \texttt{head.pt} unusable & zip cut inside the third tensor ($4{,}961{,}280$ of $5{,}248{,}703$ bytes) & complete head from the training run \\
B7 & Mac build $\neq$ published build & int4 rounding and rotary tables depend on the architecture ($\approx0.02\%$ of int4 bytes) & build on Linux x86\_64 \\
\tabalt B8 & \stwo{} deliberates for $400$ tokens without an answer; the no-think instruction is ignored & the 4-bit checkpoint does not close its deliberation block & forced \texttt{</think>} prefix (M8) \\
B9 & \stwo{} answers a generic greeting in Chinese & the assistant-turn special tokens went through the sanitiser as text and were rewritten & special tokens always by id \\
\tabalt B10 & \sone{} ``stops working'' after a patch although the served code is correct & module cached by the browser against a regenerated file & hard reload after each library change \\
B11 & generic question gives $0.040$ on the XSS (looks broken) & not a bug: the readout is conditioned on the question & ask directed questions (Figure~\ref{fig:question}) \\
\tabalt B12 & extension does not load in stable Chrome & stable Chrome ignores \texttt{-{}-load-extension} & Chrome for Testing for automation; developer mode for users \\
B13 & bare module imports fail in extension pages & import maps do not apply to extension pages & import vendored modules by URL and inject them into the runner \\
\tabalt B14 & ``both async and sync fetching of the wasm failed'' & the runtime bundle does not embed its WebAssembly & vendor the binaries ($55$\,MB); \texttt{'wasm-unsafe-eval'} in the content-security policy \\
B15 & \texttt{runtime.sendMessage} does not reach the offscreen document; \texttt{createDocument} inconsistent with \texttt{getContexts} & MV3 offscreen messaging & engine in a dedicated tab; bus over \texttt{storage.onChanged} \\
\tabalt B16 & Transformers.js hangs, without error, loading the tokenizer & \texttt{localModelPath} set to an absolute URL on another origin & embed the tokenizer in the extension (same origin); only the graph is remote \\
B17 & navigations stuck ``loading'', content script never injected & window without operating-system focus; \texttt{document\_idle} deferred indefinitely & \texttt{headless=new} (extensions and WebGPU work, no permissions) and \texttt{document\_end} \\
\tabalt B18 & \stwo{} narrates a pseudo-deliberation as literal text despite the prefix & sampling at temperature $0.7$ & directive prompt, temperature $0.3$, strip the narrated block \\
B19 & stale explanations read from storage during tests & no timestamp on the explanation record & compare the record's time with the request's \\
\bottomrule
\end{tabularx}
\end{table}

\begin{table}[ht]
\centering\bistable\footnotesize
\caption{\textbf{Firefox port and validation}, F1--F6.}
\label{tab:ffbugs}
\begin{tabularx}{\textwidth}{@{}l Y Y@{}}
\toprule
\tabhead \textbf{\#} & \textbf{Obstacle} & \textbf{Resolution} \\
\midrule
F1 & Firefox MV3 has no \texttt{background.service\_worker}; Chrome MV3 rejects \texttt{background.scripts} & two manifests, one build script producing the Firefox tree and \texttt{.xpi} \\
\tabalt F2 & no WebGPU adapter in stable Firefox or headless Nightly on macOS, also with the fallback adapter and preferences & validate in Nightly with a window (Metal adapter) \\
F3 & the driver refuses to execute script or navigate in privileged \texttt{moz-extension://} contexts & testability bridge: verdicts exposed on the page as data attributes and an event \\
\tabalt F4 & a web page cannot open \texttt{moz-extension://} URLs & the popup opens only from the browser UI or a driver-created tab \\
F5 & the \texttt{sendMessage} channel may close before a response & fire-and-forget senders; nothing depends on the synchronous reply \\
\tabalt F6 & after long headless sessions Chrome's model load hangs without error or network traffic & restart with a fresh profile; a clean regression run scans in $1.7$\,s \\
\bottomrule
\end{tabularx}
\end{table}

\end{document}